\documentclass[]{aastex7}

\begin{document}

\title{Characterizing the Nature of Periodic Amplitude Modulation in Pulsars}

\author[0000-0003-1824-4487]{Rahul Basu}
\affiliation{Janusz Gil Institute of Astronomy, University of Zielona G\'ora, ul. Szafrana 2, 65-516 Zielona G\'ora, Poland.}
\email{rahulbasu.astro@gmail.com}

\author[0000-0002-9142-9835]{Dipanjan Mitra}
\affiliation{National Centre for Radio Astrophysics, Tata Institute of Fundamental Research, Pune 411007, India.}
\affiliation{Janusz Gil Institute of Astronomy, University of Zielona G\'ora, ul. Szafrana 2, 65-516 Zielona G\'ora, Poland.}
\email{dmitra@ncra.tifr.res.in}

\author[0000-0003-1879-1659]{George I. Melikidze}
\affiliation{Janusz Gil Institute of Astronomy, University of Zielona G\'ora, ul. Szafrana 2, 65-516 Zielona G\'ora, Poland.}
\affiliation{Evgeni Kharadze Georgian National Astrophysical Observatory, 0301 Abastumani, Georgia.}
\email{G.Melikidze@ia.uz.zgora.pl}

\begin{abstract}
In recent years periodic amplitude modulation has emerged as a unique emission
feature in the single pulse sequence of pulsars alongside periodic nulling and 
subpulse drifting. Despite ample evidence for the uniqueness of this 
phenomenon, the periodic modulation in several pulsars are often confused with 
subpulse drifting, primarily due to lack of clear characterisation of the 
emission features from a representative sample of pulsars. In this work we 
present a detailed analysis of the single pulse behaviour from seventeen 
pulsars exhibiting periodic amplitude modulation, six of them being new 
detections. The pulsar switches between different intensity states as a result 
of periodic amplitude modulation and we propose a novel statistical scheme to 
identify these emission states. The periodic modulation can be divided into 
three broad categories, phase stationary modulation, modulations with phase 
shift and intermittent periodic modulations. The phase stationary behaviour is 
seen when the emission intensity across a major part of the pulse window 
changes periodically. The phase shifts are associated with intensity changes at
specific locations within the emission window in a periodic manner; while in 
some pulsars the periodic modulations become more prominent only at specific 
intervals resulting in intermittent behaviour. 
\end{abstract}


\section{Introduction} 
The periodic modulations in the single pulse sequence of pulsars have been 
detected soon after their discovery, in the form of subpulse drifting 
\citep{DC68}, that appear as systematic drift bands across the emission window.
Subpulse drifting has emerged as one of the main observational features of 
pulsar radio emission, with more than hundred sources showing this behaviour 
\citep{WES06,BMM16}, and the phenomenon serving as a key diagnostic of the 
plasma generation process in the inner acceleration region \citep{RS75,GMG03,
BMM22b,BMM23a}. Detailed studies have revealed several distinguishing aspects 
of the drifting behaviour; drifting is only seen in the conal components of the
pulsar profile and is absent in the central core region \citep{R86,BMM19}; the 
phenomenon is restricted to the lower energetic pulsars, spin-down energy loss 
$\dot{E} <  5\times10^{32}$ ergs~s$^{-1}$, and the drifting periodicity is 
anti-correlated with $\dot{E}$ \citep{BMM16,BMM19}. An increasing population of
pulsars shows periodic modulations which do not resemble the prominent drift 
bands associated with subpulse drifting. Some of these can be attributed to 
specific line of sight traverse across the emission beam, particularly central 
cuts with multiple component profiles like PSRs B0844$-$35 \citep{BMM23b}, 
B1237+25 \citep{MD14}, J1239+0326 \citep{RZW25}, B1737+13 \citep{FR10}, 
B1758$-$29 \citep{BMM23b}, B1857$-$26 \citep{MR08}, B2000+40 \citep{BLK20}, 
etc. The fluctuation spectral analysis \citep{B73} shows the presence of 
drifting features with relatively flat phase variations only in the conal 
regions and a lack of any drifting features in the core component. 

The other periodic behaviour seen in the pulsar single pulse sequence include 
periodic nulling and periodic amplitude modulation that appear to be quite 
distinct from subpulse drifting. The periodic nulling is seen when the pulsar
emission goes below detection level at regular intervals across the entire
pulsed window. In some cases the periodic nulling co-exists with subpulse 
drifting, which initially led to the suggestion that they represent empty line
of sight traverse between the drifting pattern \citep{HR07,HR09}. But 
subsequent works have shown that periodic nulling has very different physical
properties compared to subpulse drifting. They appear as low frequency features
in the fluctuation spectra across both the core and conal components, and are 
also seen in more energetic pulsars, thereby suggesting a different physical 
phenomenon being responsible for periodic nulling \citep{BMM17,BMM20a}. On the
other hand in the case of periodic amplitude modulation the intensity within 
the emission window changes periodically without the presence of any systematic
drift bands which led to their separate categorisation \citep{BMM16,BMM20a}. 
Similar to periodic nulling the periodic amplitude modulations are also seen in
both core and conal emission and over a much wider $\dot{E}$ range of pulsars.

Detailed studies of the periodic amplitude modulation behaviour in several 
individual sources have been conducted since the initial identification of this 
phenomenon in pulsars. In PSR B1946+35, a triple profile pulsar comprising of a
central cone surrounded by a conal pair, the modulation show intermittent 
behaviour with the transitions between emission states showing high levels of 
periodicity at short intervals lasting several hundred pulses \citep{MR17}. The
intensity variation is not uniform across the profile, where the leading cone 
and core are in the higher intensity state when the trailing component is 
weaker and vice versa \citep[also see][]{CWK25}. Several other sources also 
exhibit different forms of intensity variations across the emission window in 
the two states of periodic amplitude modulation. PSR B0823+26 shows this 
phenomenon during its B mode, with the emission window becoming wider during 
the bright state \citep{BM19}. The periodic modulation is seen only in the 
leading component of PSR J1722$-$3207 with a double profile \citep{ZYW23}, 
while there is also an indication that such modulations are present at the 
trailing side of the Vela pulsar \citep{WCH20}. The periodic modulations have 
been detected in both the main pulse and interpulse regions of pulsars with 
identical periodicities, e.g. PSRs B0823+26 \citep{CWD23}, Q mode of B1822$-$09
\citep{LMR12,YMW19} and B1929+10 \citep{KYP21}, suggesting that the mechanism 
responsible for these modulations may not be localized to the polar cap region.
It has also been possible to separate the two intensity states during periodic 
amplitude modulation using statistical estimates, in PSRs J1048$-$5832 
\citep{YMW20}, J1722$-$3207 \citep{ZYW23}, J1921+1419 \citep{TXD25}, etc. The 
average profiles of the two emission states show that the polarization position
angle traverse remains largely identical, suggesting that the radio emission 
from each state originate from similar locations within the open field line 
region of the pulsar \citep[see discussion in][]{MBM24a}.

The periodic amplitude modulation has also been seen in different types of
sources, like the young energetic Vela pulsar \citep{WCH20}, the millisecond 
pulsar J0621+1002 \citep{WWW21} and in the intermittent pulsar J1841$-$0500 
\citep{WWH20}. The co-existence of subpulse drifting and periodic nulling in 
the single pulse sequence is a fairly common occurrence with more than ten 
sources reported at present \citep{BMM20a}. But the periodic amplitude 
modulation and subpulse drifting happen less often in the same system, with 
only two reported cases, PSRs B1737+13 \citep{FR10} and B2021+51 \citep{CWW24}.
Recently, the Thousand Pulsar Array programme has significantly increased the 
number of pulsars with periodic modulations, where more than five hundred 
sources have been reported to exhibit some form of periodic feature in their 
fluctuation spectra \citep{SWS23}. While the single pulse emission behaviour 
corresponding to subpulse drifting and periodic nulling have been explored in 
some detail, characterisation of the single pulse behaviour of periodic 
amplitude modulation in a representative sample of pulsars is somewhat lacking,
apart from the few individual sources mentioned above. We have investigated the
single pulse properties of seventeen pulsars showing periodic amplitude 
modulation, with six cases being reported for the first time, in order to 
characterise their radio emission properties. In section \ref{sec:analysis} we 
explain the analysis process where a statistical scheme for separating the two 
emission states has been introduced. The emission behaviour from each pulsar 
have been described in section \ref{sec:Per_Mod}, and we find that the periodic
amplitude modulations can be divided into three distinct categories, the more 
conventional phase stationary modulation reported in section 
\ref{sec:Phs_Stat}, the modulations with phase shifts in section 
\ref{sec:Phs_Sft} and intermittent modulations reported in section 
\ref{sec:Int_Var}. The implications of the periodic modulation categories and 
possible ways to distinguish them from subpulse drifting are discussed further 
in section \ref{sec:Discuss}.

\section{Observation and Analysis} \label{sec:analysis}

\centerwidetable
\begin{deluxetable}{ccccccccccc}
\tablecaption{Observational Details
\label{tab:srclist}}
\tabletypesize{\small}
\tablewidth{0pt}
\tablehead{
 \colhead{Name} & \colhead{Frequency} & \colhead{Class} & \colhead{$P$} & \colhead{$\dot{E}$} & \colhead{Npulse} & \colhead{SNR} & \colhead{$W_{5\sigma}$} & \colhead{$W_{10}$} & \colhead{$P_M$} & \colhead{Reference} \\
   & \colhead{(MHz)} &   & \colhead{(sec)} & \colhead{(erg s$^{-1}$)} &   &   & \colhead{(deg)} & \colhead{(deg)} & \colhead{($P$)} &  }
\startdata
  B0136+57  & 300-500 & S$_t$ & 0.272 & 2.09$\times10^{34}$ & ~833 & 150.2 & 18.9$\pm$0.9 & 10.7$\pm$0.9 & 8.4$\pm$0.8 & [1] \\
 B0450$-$18 & 306-339 & T & 0.549 & 1.38$\times10^{33}$ & 2730 & 177.0 & 39.5$\pm$0.7 & 26.2$\pm$0.7 & 16$\pm$6 & [2] \\
  B0450+55  & 306-339 & T & 0.341 & 2.37$\times10^{33}$ & 2655 & 178.4 & 40.5$\pm$1.1 & 30.7$\pm$1.1 & 9$\pm$1 & [2] \\
 B0905$-$51 & 300-500 & T & 0.254 & 4.44$\times10^{33}$ & 3541 & 137.7 & 135.9$\pm$0.9 & 99.2$\pm$0.9 & 52$\pm$27 & --- \\
  B1541+09  & 306-339 & T & 0.748 & 4.09$\times10^{31}$ & 2165 & 54.5 & 173.0$\pm$0.5 & --- & 15$\pm$5 & [2] \\
 B1600$-$49 & 602-618 & T$_{1/2}$ & 0.327 & 1.14$\times10^{33}$ & 2197 & 35.8 & 18.2$\pm$0.5 & --- & 50$\pm$26 & [3] \\
 B1604$-$00 & 306-339 & T & 0.422 & 1.14$\times10^{33}$ & 3552 & 166.0 & 35.0$\pm$0.4 & 16.5$\pm$0.4 & 34$\pm$13 & [2] \\
 B1642$-$03 & 602-618 & S$_t$ & 0.388 & 1.20$\times10^{33}$ & 2154 & 741.3 & 22.8$\pm$0.5 & 6.8$\pm$0.5 & 13$\pm$4 & [3] \\
 B1730$-$37  & 550-750 & D & 0.338 & 1.54$\times10^{34}$ & 2471 & 15.3 & 79.7$\pm$0.7 & --- & 71$\pm$20 & [1] \\
 B1732$-$07 & 317-333 & T & 0.419 & 6.51$\times10^{32}$ & 2148 & 88.7 & 22.4$\pm$0.4 & --- & 20$\pm$8 & [3] \\
 B1737$-$39 & 602-618 & S$_t$ & 0.512 & 4.88$\times10^{32}$ & 2108 & 44.2 & 21.3$\pm$0.3 & 16.1$\pm$0.3 & 73$\pm$44 & [3] \\
  B1917+00  & 317-333 & T & 1.272 & 1.47$\times10^{32}$ & 1678 & 62.5 & 13.8$\pm$0.4 & --- & 11$\pm$0.4 & [3] \\
  B1929+10  & 300-500 & D/T$_{1/2}$ & 0.227 & 3.93$\times10^{33}$ & 2602 & 342.6 & 201.0$\pm$1.0 & 22.8$\pm$1.0 & 12$\pm$1 & --- \\
  B1929+20  & 550-750 & S$_t$ & 0.268 & 8.63$\times10^{33}$ & 2437 & 82.5 & 28.4$\pm$0.9 & 13.9$\pm$0.9 & 16$\pm$6 & [1] \\
  B2011+38  & 550-750 & S$_t$ & 0.230 & 2.86$\times10^{34}$ & 2056 & 51.3 & 63.5$\pm$1.0 & 46.6$\pm$1.0 & 27$\pm$11 & [1] \\
  B2148+52  & 550-750 & D/$_c$T & 0.332 & 1.09$\times10^{34}$ & 1772 & 70.8 & 27.1$\pm$0.7 & 19.1$\pm$0.7 & 8.8$\pm$0.5 & [1] \\
  B2334+61  & 550-750 & S$_t$ & 0.495 & 6.28$\times10^{34}$ & 1817 & 50.8 & 30.1$\pm$0.5 & 24.7$\pm$0.5 & 18$\pm$3 & [1] \\
\tableline
\enddata
\tablecomments{[1] \citet{MBM25}; [2] \citet{BMM20a}; [3] \citet{MBM16}}
\end{deluxetable}

Table \ref{tab:srclist} lists the properties of seventeen pulsars observed with
the Giant Metrewave Radio Telescope (GMRT). The GMRT comprises of 30 antennas, 
with 45 meter diameter each, spread in a Y-shaped array with maximum separation
of 25 km \citep{SAK91}. The single pulse emission from pulsars are usually 
observed in the phased-array mode where the measured intensities from around 20
nearby antennas are co-added after adjusting for their phase lags to increase 
detection sensitivity. Over the years the GMRT receiver system has undergone 
several modifications, from a 33 MHz hardware system (GHB) to a software (GSB) 
receiver backend \citep{RGP10}, and the latest upgrade to the wideband (GWB) 
receiver system with 200 MHz bandwidth \citep{GAK17}. This work primarily uses 
archival observations, covering more than a decade of pulsar studies with GMRT 
and have been reported in previous works (see Table \ref{tab:srclist}, last 
column). The table shows the pulsar name, the observing frequency range, the 
profile classification, the rotation period ($P$) and $\dot{E}$ obtained from 
ATNF catalog \citep{MHT05}. We limited our studies to pulsars with high 
detection sensitivity of single pulses, and the signal to noise ratio (SNR) of 
the pulsed emission in the average profile is listed in Table 
\ref{tab:srclist}. The table also lists the number of pulses during the 
observing duration and the widths of the average profiles measured at the 
profile edges, with intensities of five standard deviations above the mean 
baseline level ($W_{5\sigma}$), and at 10\% level of the peaks of the outer 
most components ($W_{10}$). The details of the observing setup for the 
narrowband measurements are available in \citet{BMM16,BMM19,BMM20a}, where the 
single pulse emission from more than 200 pulsars have been studied. More 
recently \citet{MBM25} used the wideband system to investigate the single pulse
polarization behaviour of around 40 high energetic pulsars with $\dot{E} > 
5\times10^{33}$ ergs~s$^{-1}$. We explored the single pulse properties of this 
sample and detected periodic amplitude modulation for the first time in six 
pulsars. Additionally, PSRs B0905$-$51 and B1929+10 were observed in December 
2018 as part of a separate observational project, using the 300-500 MHz 
frequency band of GWB. 

\begin{figure}
\gridline{\fig{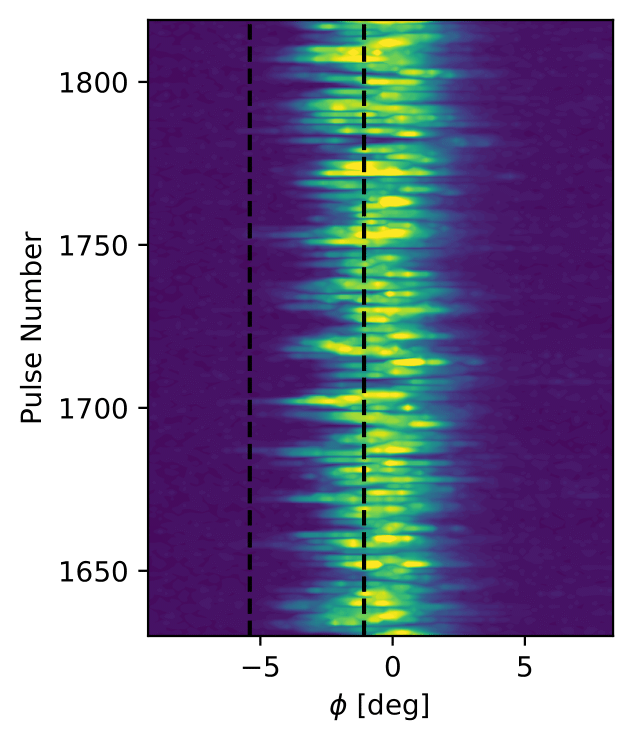}{0.38\textwidth}{(a)}
          \fig{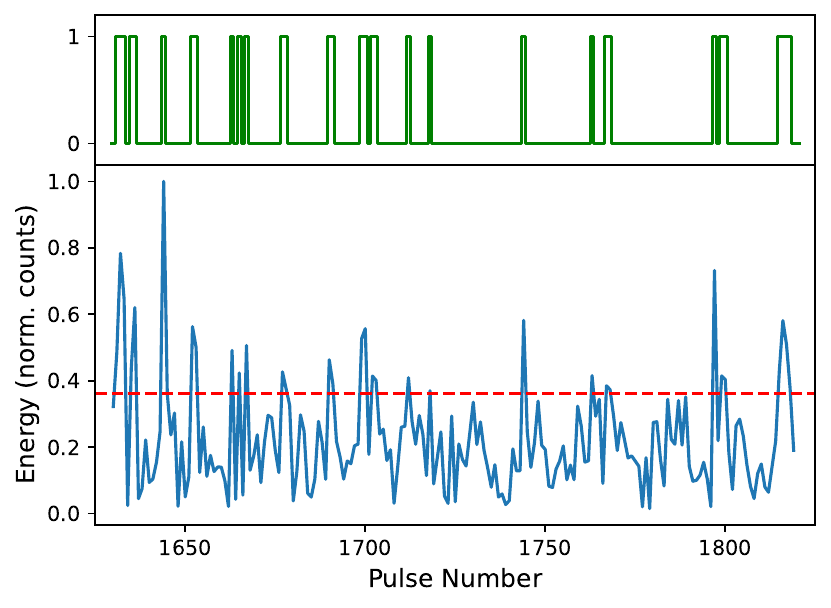}{0.42\textwidth}{(b)}
         }
\caption{(a) A short single pulse sequence of PSR B1642-03 with periodic 
amplitude modulation. (b) The lower window shows the average intensity of each 
pulse estimated within the longitude range specified by the two vertical dashed
lines in the left panel. A statistical cutoff level for the two emission states
is shown as the horizontal, red-dashed line and the top window shows the 0/1 
time sequence obtained using this cutoff.
\label{fig:B1642_singl}}
\end{figure}

\begin{figure}
\gridline{\fig{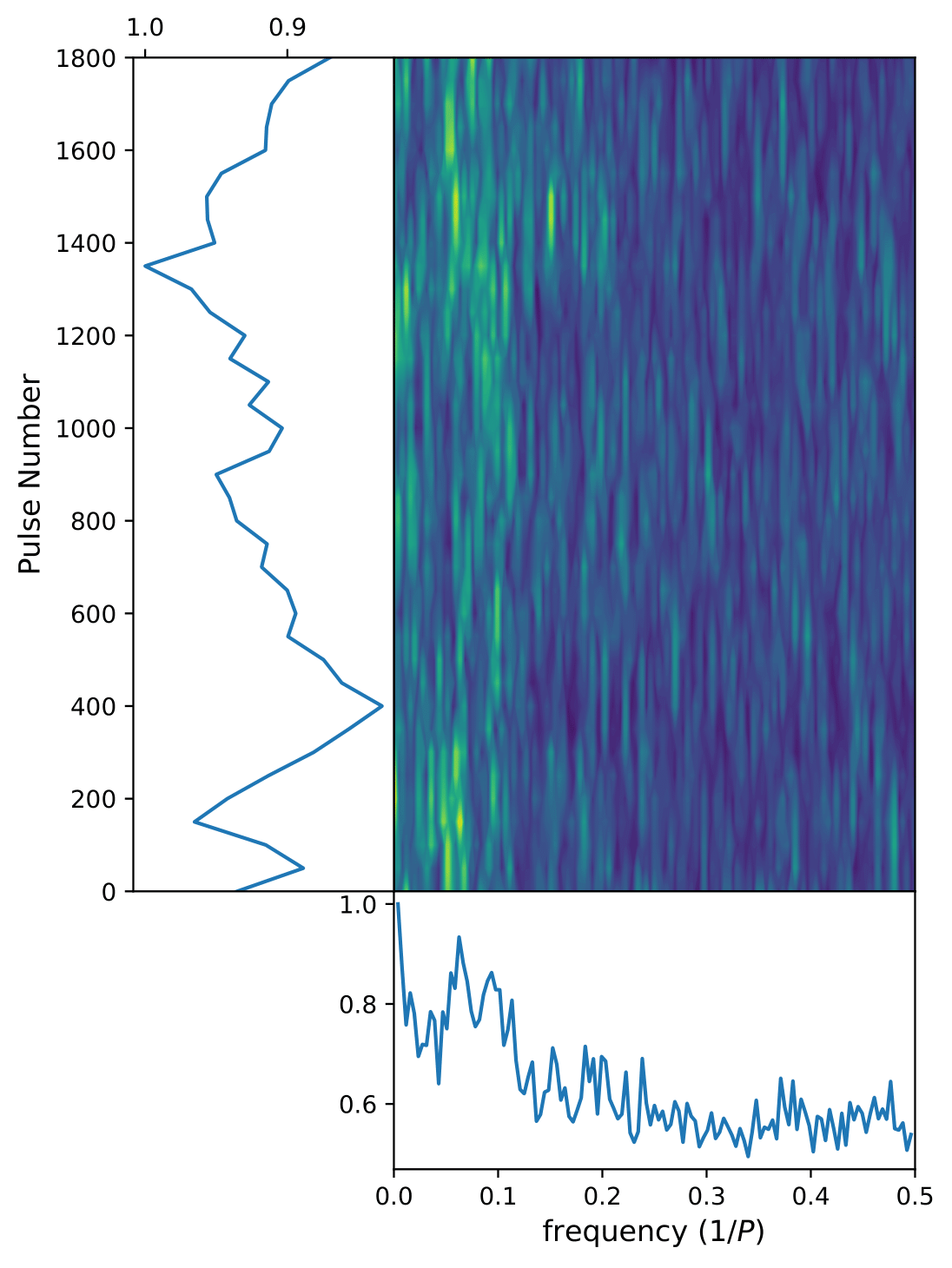}{0.38\textwidth}{(a) Time averaged LRFS}
          \fig{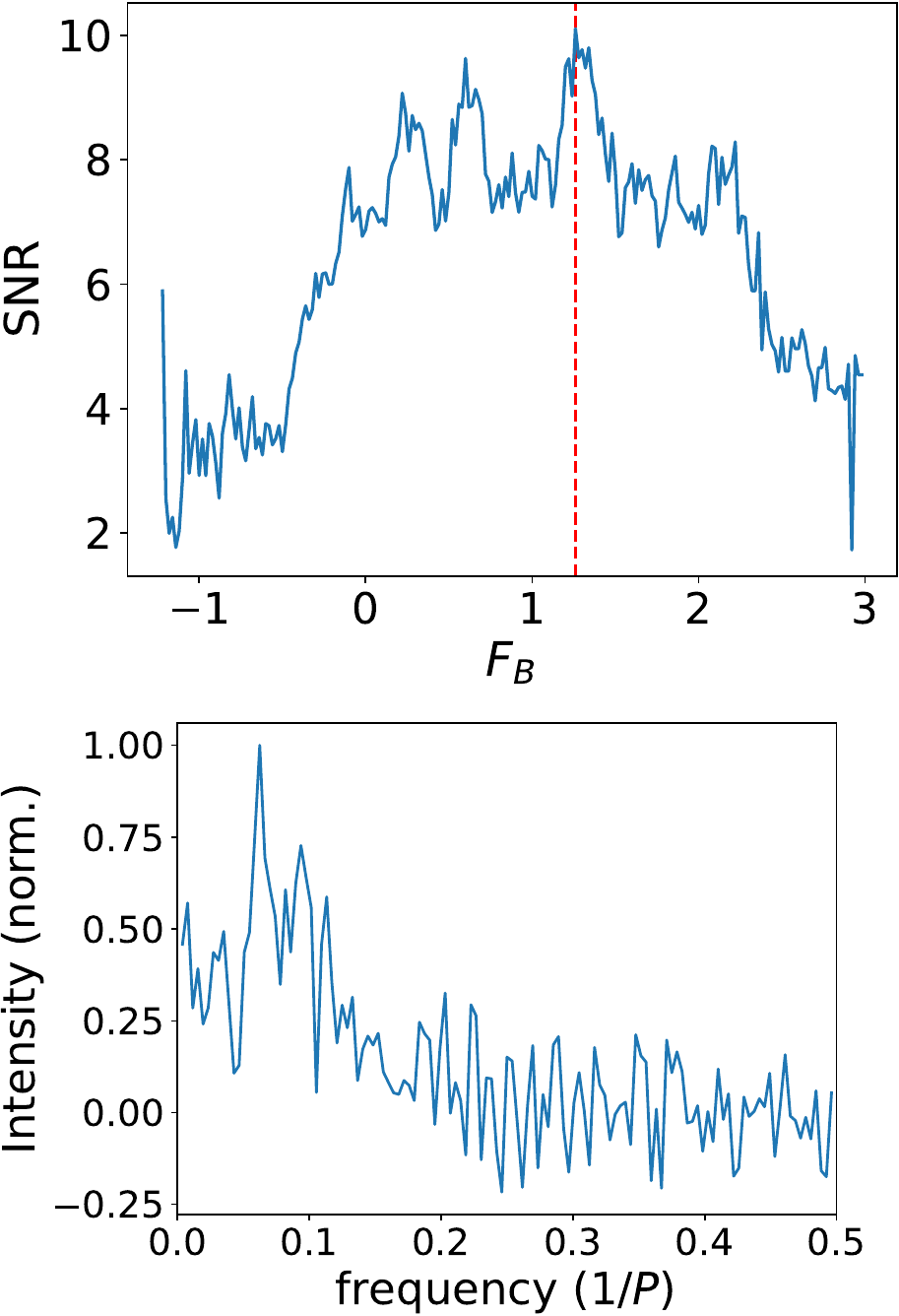}{0.34\textwidth}{(b) Time averaged FFT}
         }
\caption{(a) The time varying longitude-resolved fluctuation spectra (LRFS) 
estimated on the single pulse sequence of PSR B1642-03, showing the broad 
periodic amplitude modulation feature between frequency range 0.05--0.1 
cycles/$P$. (b) The 0/1 time series FFT is estimated for different cutoff 
levels and the signal to noise ratio (SNR) of the periodic feature is shown in 
the top window. The maximum SNR corresponds to 1.2 times the standard deviation
above the median level (dashed vertical red line). The bottom window shows the 
average FFT of the 0/1 sequence with maximum SNR of periodic feature.
\label{fig:B1642_permod}}
\end{figure}

The periodic amplitude modulation is seen when the single pulse emission 
has different intensity states and the pulsar transitions between them in a 
periodic/quasi-periodic manner. The periodicity of the modulation ($P_M$) can 
be measured using the longitude-resolved fluctuation spectra 
\citep[LRFS,][]{B73} and the time varying average LRFS \citep[see][for more 
details]{BMM16,BM18a}, and Table \ref{tab:srclist} lists the estimated 
periodicity in each pulsar. However, it is more challenging to separate the two
intensity states and estimate their emission properties, and only in a few 
cases like PSRs J1048$-$5832 \citep{YMW20}, J1722$-$3207 \citep{ZYW23} and 
J1921+1419 \citep{TXD25} there are clear statistical boundaries between the 
intensities of the two states for such studies to be possible. We propose an 
iterative scheme for identifying the two emission states associated with 
periodic amplitude modulation. The single pulse energy distribution is 
estimated and an initial on-off sequence is setup by putting all pulses with 
energies ($S$) above a certain cutoff factor ($\cal{F}$) times standard 
deviation ($\sigma$) from the median value ($\cal{M}$), i.e.  $S > \cal{M} + 
\cal{F}\sigma$, as `1' and the remaining pulses as `0'. A time varying FFT is 
carried out similar to the periodic nulling analysis \citep[see][for 
details]{BMM17}. The SNR of the periodic feature in the average FFT spectra is 
estimated and the process is repeated with a different $\cal{F}$ till the 
highest sensitivity is reached, which corresponds to $\cal{F}=\cal{F}_{\rm B}$.
The on and off states in this scenario largely represent the two intensity 
states of periodic amplitude modulation.

In pulsars where the intensity variations are more pronounced at specific 
locations within the pulsed window a narrower longitude range can be used for 
estimating the intensity levels to better separate the two emission states. 
Fig.~\ref{fig:B1642_singl}(a) shows a short single pulse sequence of PSR 
B1642-03 where the periodic amplitude modulation is seen more prominently near 
the leading side of the emission window and the intensity distribution of the 
single pulses is estimated from this region (dashed vertical lines). The time 
varying average LRFS of PSR B1642-03 is shown in Fig.~\ref{fig:B1642_permod}(a) 
and a wide, double-peaked, low frequency feature is clearly visible. The on-off
sequence for a specific value of $\cal{F}$ is obtained (see 
Fig.~\ref{fig:B1642_singl}b) and the SNR of the periodic feature in the FFT is 
estimated. The value of $\cal{F}$ is changed to gradually shift the cutoff 
level in the vertical direction and the variation of SNR is shown in 
Fig.~\ref{fig:B1642_permod}(b) (top window). The maximum SNR of the periodic 
feature is seen when $\cal{F}_{\rm B}$ = 1.2, which serves as the statistical 
boundary between the intensities of the two states. The FFT of the 0/1 sequence
in Fig.~\ref{fig:B1642_permod}(b) (bottom window) has periodic behaviour that 
closely resembles the low frequency feature of periodic amplitude modulation in
the average LRFS.

\section{Periodic Amplitude Modulation: Emission features} \label{sec:Per_Mod}

There are primarily three different types of periodic behaviour seen in our 
sample. In seven pulsars the intensity variations take place either across the 
full profile or at specific locations without significant shifts of the 
emission window, and this behaviour is classified as phase stationary 
modulation. In the second group containing six pulsars, the two emission states
are seen across different longitudinal ranges with shifted emission windows and
this has been named as periodic modulations with phase shifts. In the remaining
four pulsars the periodic behaviour is seen sporadically, notable only at 
specific time intervals, with low SNR feature in the average LRFS. The 
detection sensitivity of the single pulses in these four sources was comparable
to pulsars in the previous two categories (see Fig.~\ref{fig:state_singl}), and
this phenomenon has been classified as intermittent periodic modulation. 

In the thirteen pulsars belonging to the first two categories it was possible 
to separate the two emission states of periodic amplitude modulation using the
statistical scheme described above. Table \ref{tab:ampstate} reports the 
estimated $\cal{F}_{\rm B}$ boundary between the two states of each pulsar, and
single pulses with intensities above this level belong to the `Bright' emission
state and remaining are part of the `Weak' emission state. The table also lists
the fractional abundance of each emission state during the observing duration. 
The average profiles of the two states were estimated separately to further 
investigate their emission properties (see Fig.~\ref{fig:state_prof}). The 
profile widths of the two emission states as well as the ratios between their 
peak and average intensities have been estimated. Below we describe in detail
the emission properties of the individual sources belonging to the three types
of periodic amplitude modulation.

\subsection{Phase Stationary Modulation} \label{sec:Phs_Stat}

\begin{figure}
\gridline{\fig{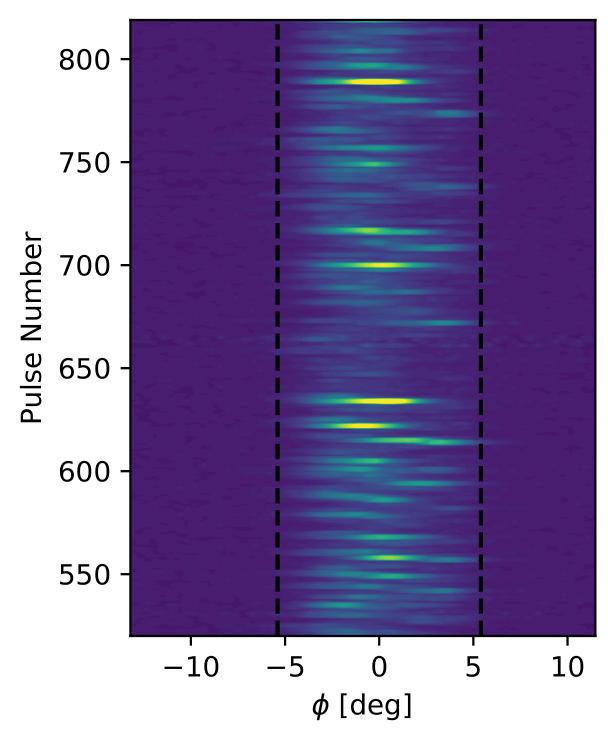}{0.215\textwidth}{(a) PSR B0136+57}
	  \fig{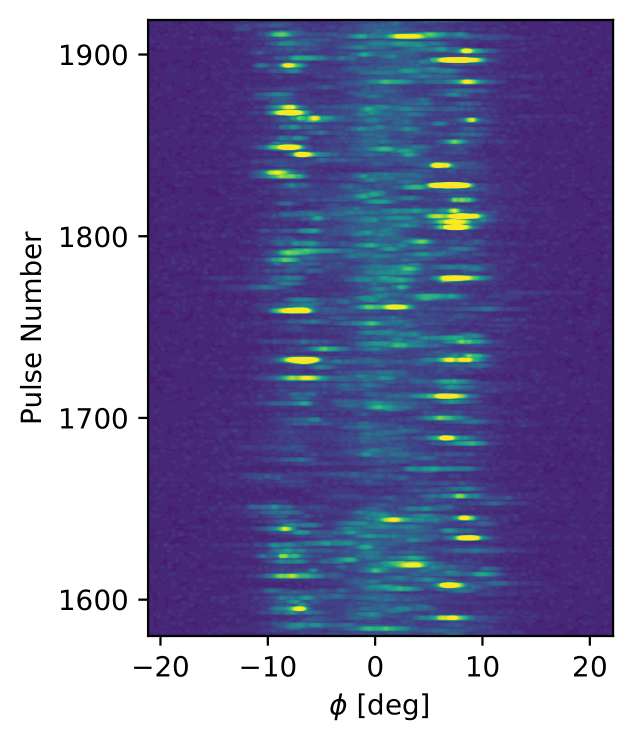}{0.22\textwidth}{(b) PSR B0450$-$18}
	  \fig{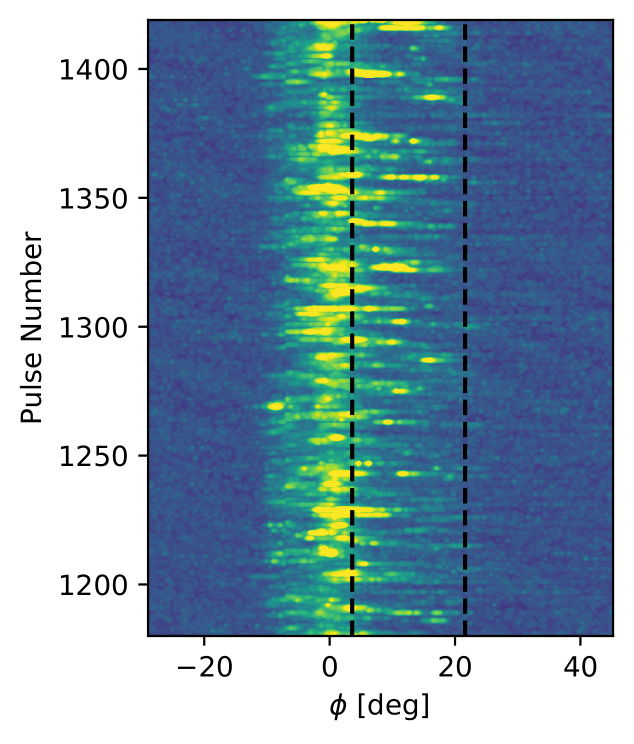}{0.22\textwidth}{(c) PSR B0450+55}
	  \fig{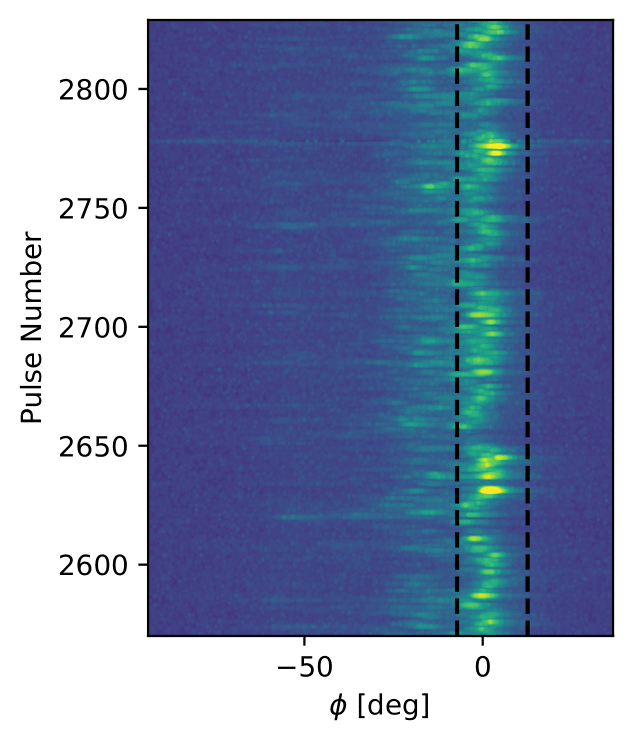}{0.22\textwidth}{(d) PSR B0905$-$51}
         }
\gridline{\fig{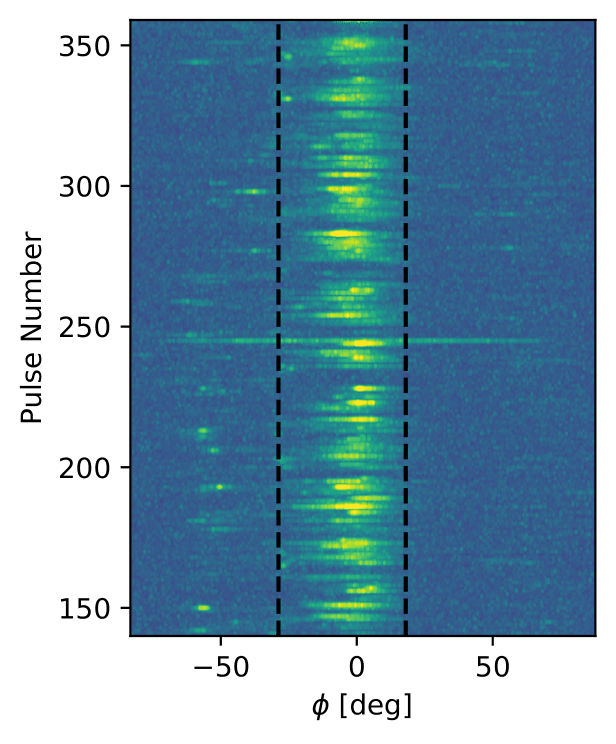}{0.21\textwidth}{(e) PSR B1541+09}
	  \fig{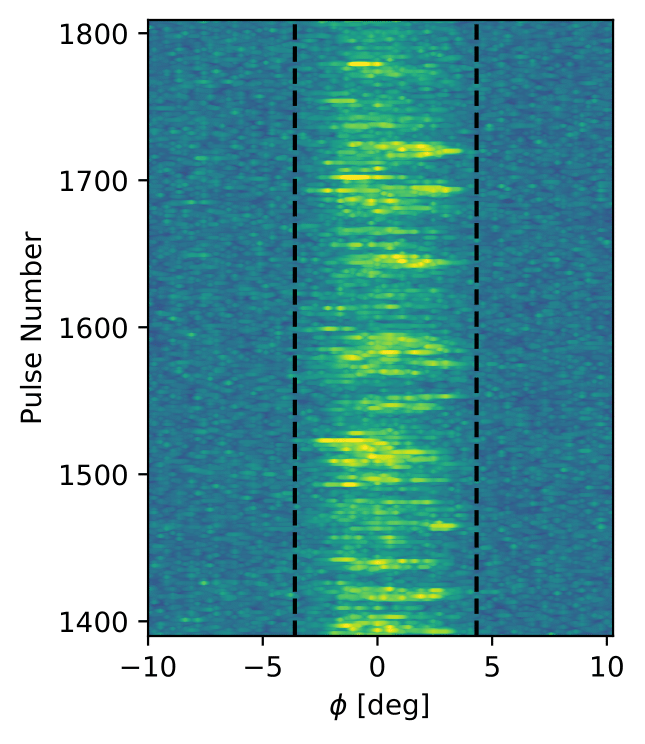}{0.22\textwidth}{(f) PSR B1600$-$49}
	  \fig{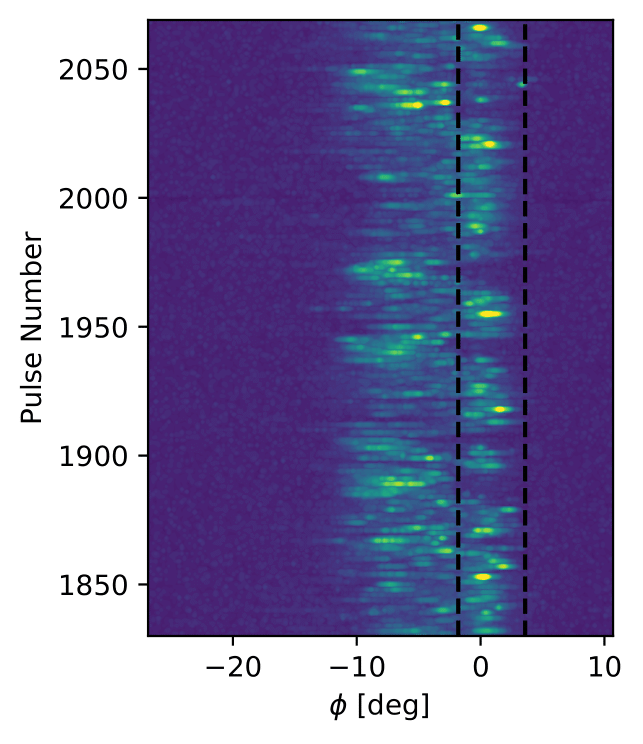}{0.22\textwidth}{(g) PSR B1604$-$00}
	  \fig{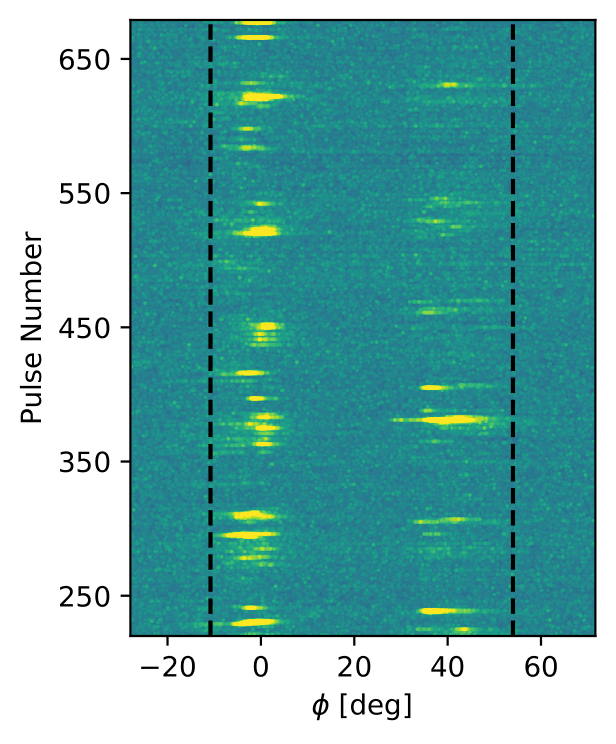}{0.215\textwidth}{(h) PSR B1730$-$37}
	 }
\gridline{\fig{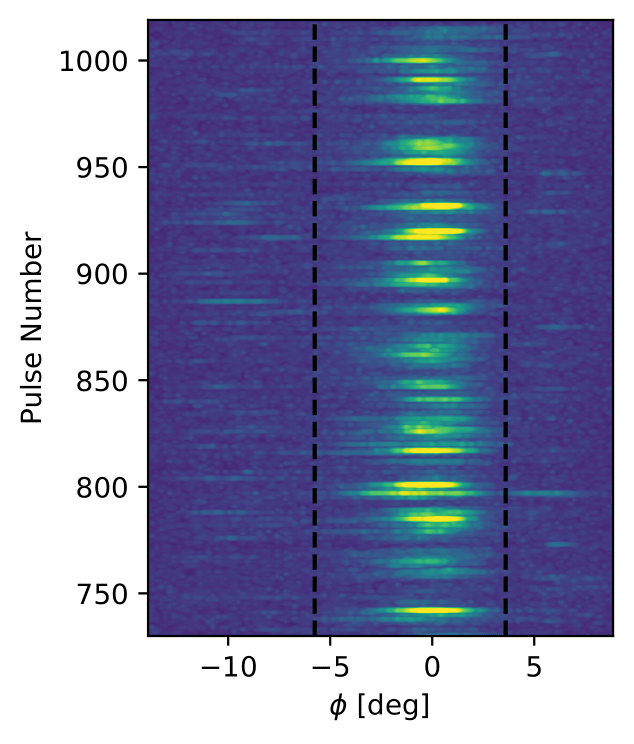}{0.225\textwidth}{(i) PSR B1732$-$07}
	  \fig{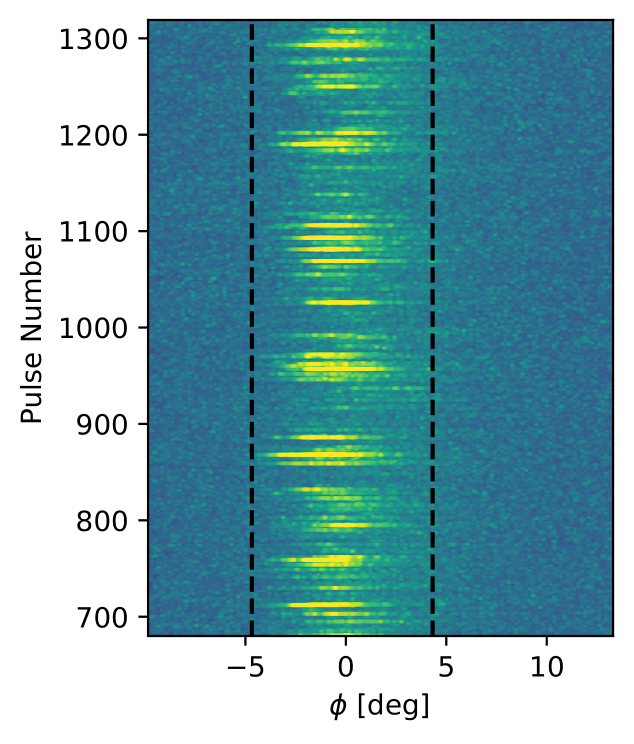}{0.225\textwidth}{(j) PSR B1737$-$39}
	  \fig{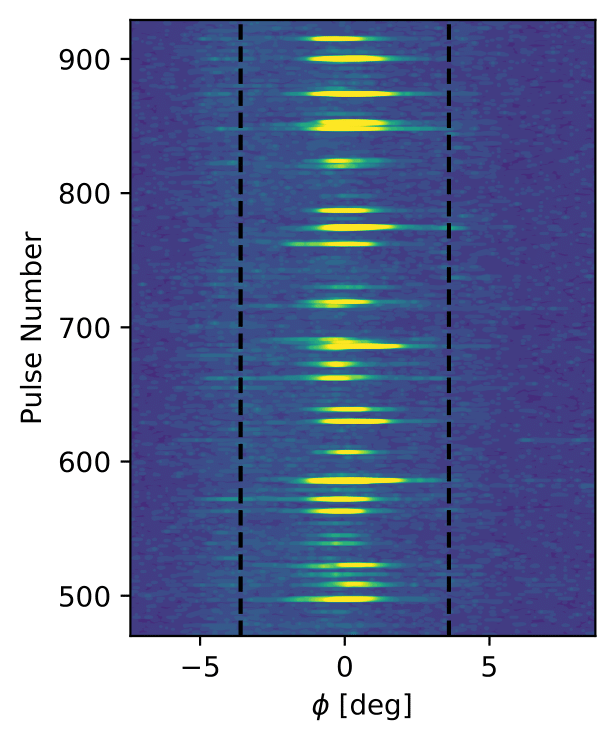}{0.22\textwidth}{(k) PSR B1917+00}
	  \fig{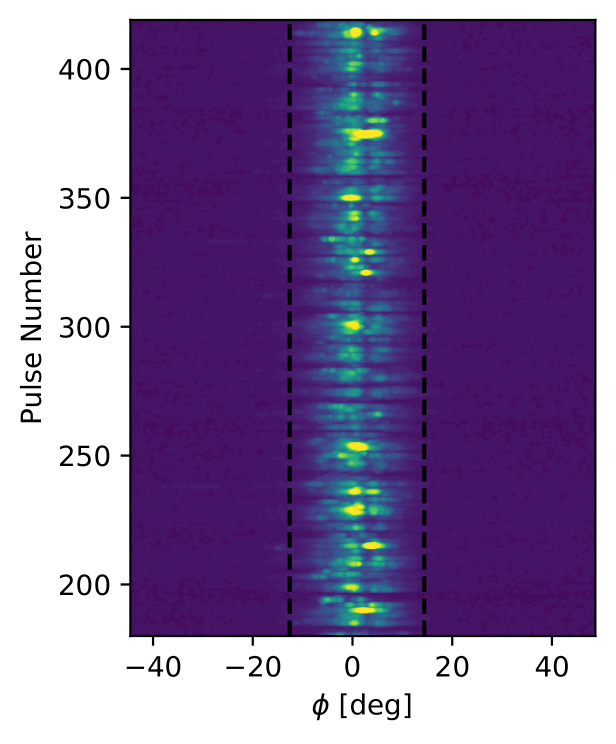}{0.22\textwidth}{(l) PSR B1929+10}
	 }
\gridline{\fig{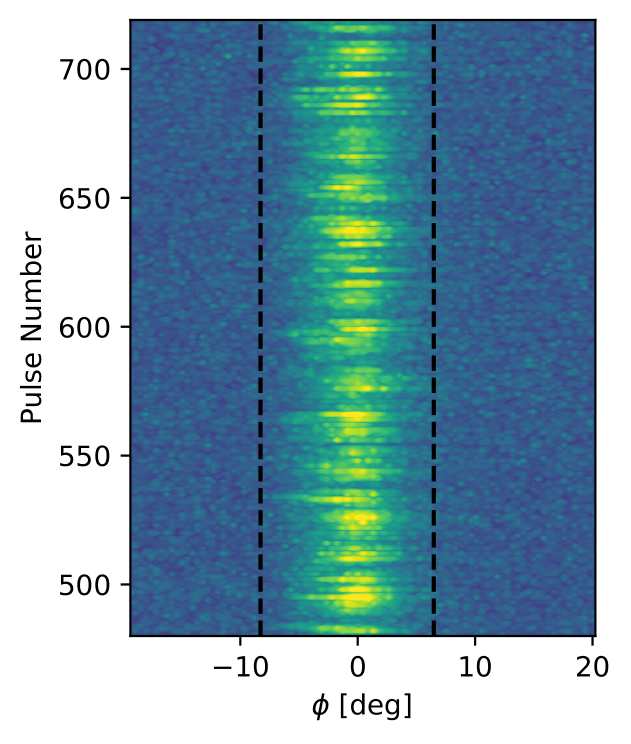}{0.225\textwidth}{(m) PSR B1929+20}
	  \fig{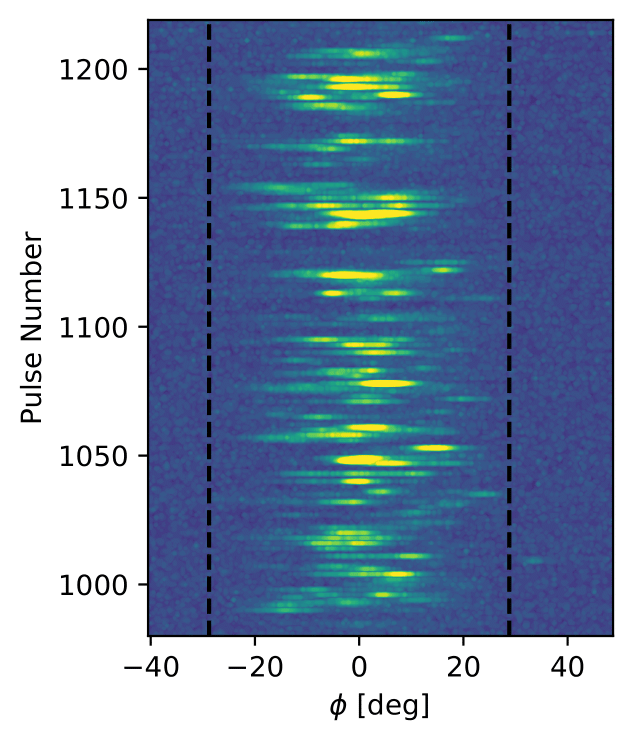}{0.225\textwidth}{(n) PSR B2011+38}
	  \fig{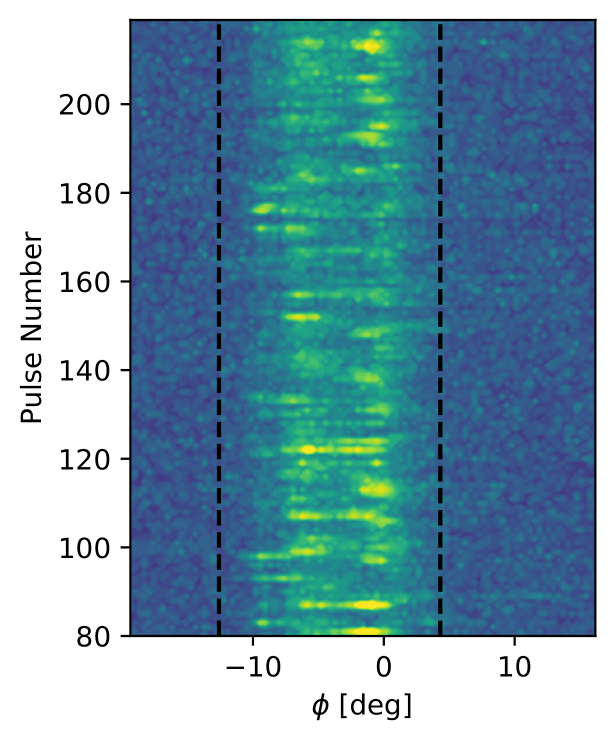}{0.22\textwidth}{(o) PSR B2148+52}
	  \fig{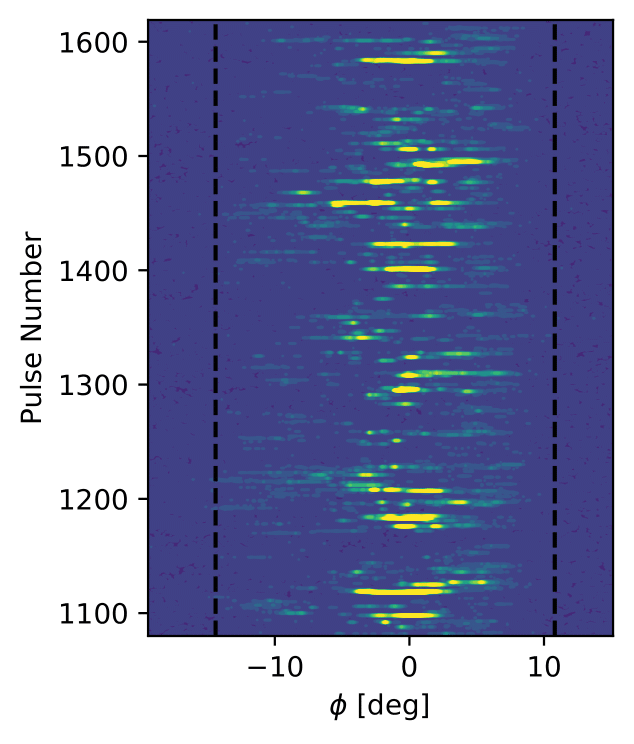}{0.22\textwidth}{(p) PSR B2334+61}
	 }
\caption{Single pulse sequence showing the nature of the periodic amplitude 
modulation in each pulsar. The dotted lines show the longitude window within
which the average intensities are estimated for the state separation studies.
\label{fig:state_singl}}
\end{figure}

\begin{figure}
\gridline{\fig{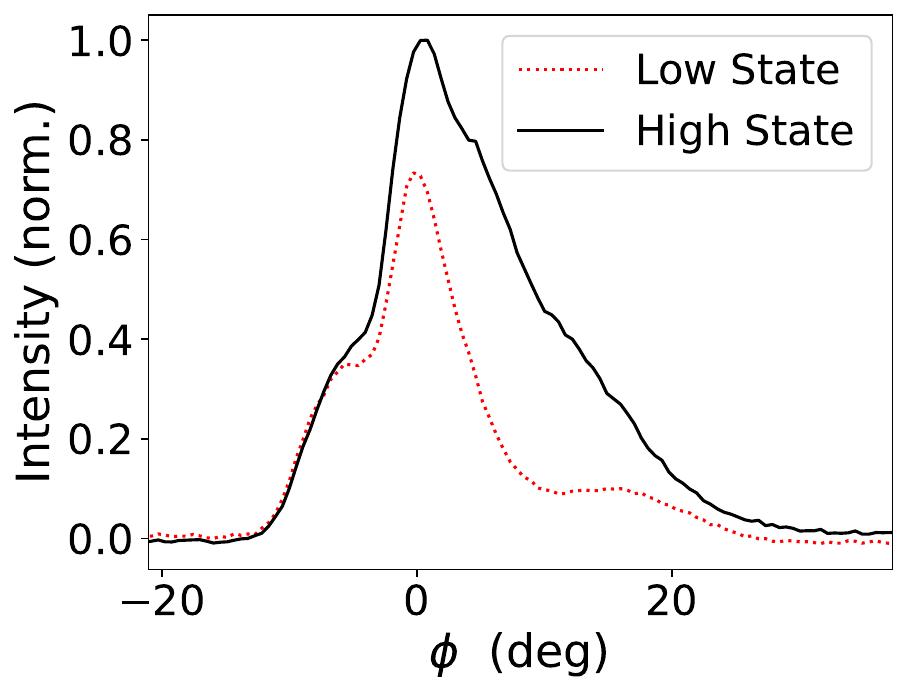}{0.23\textwidth}{(a) PSR B0450+55}
	  \fig{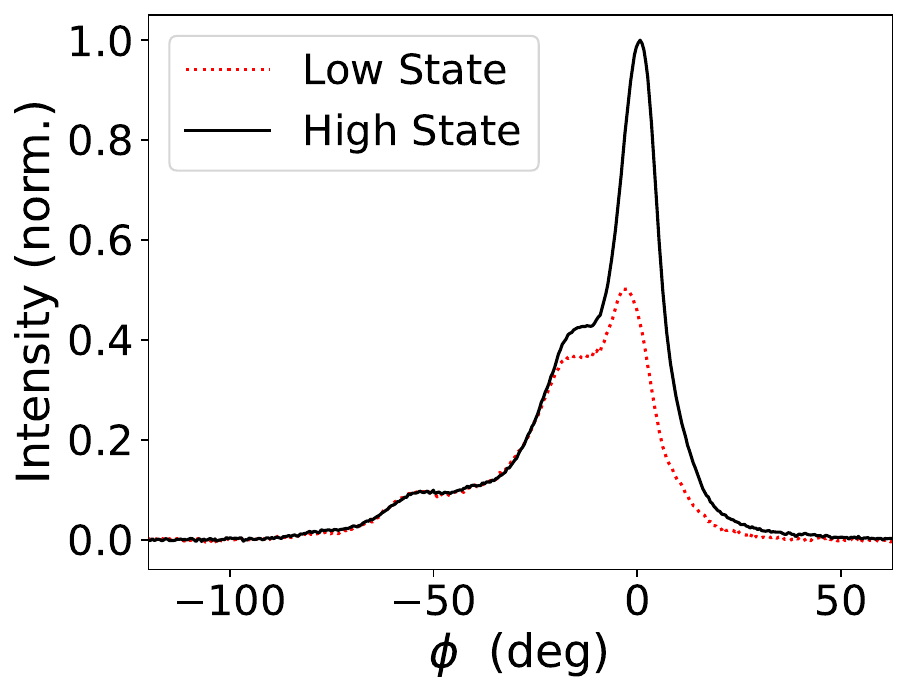}{0.23\textwidth}{(b) PSR B0905$-$51}
	  \fig{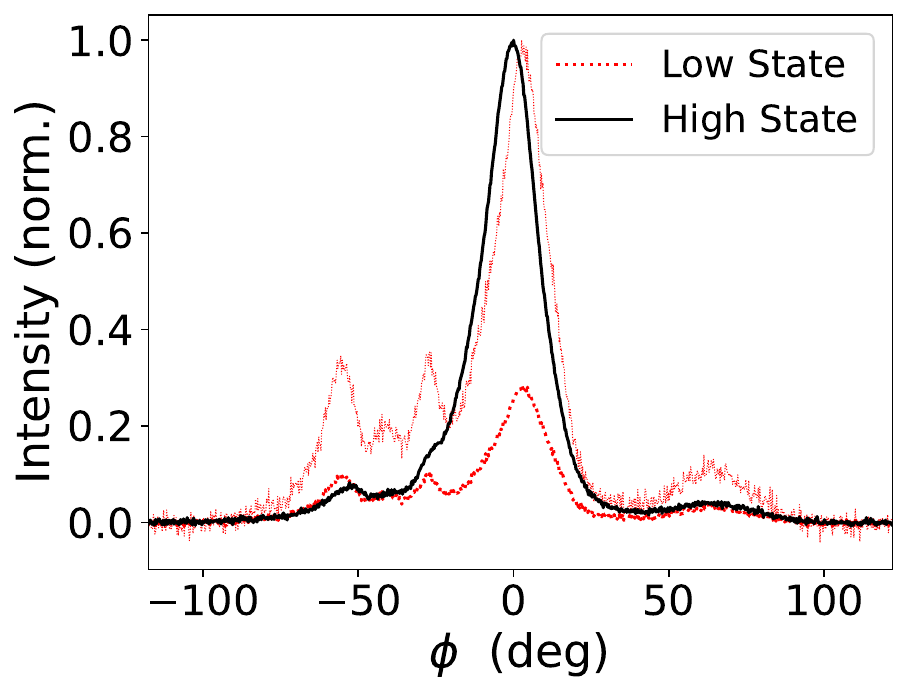}{0.23\textwidth}{(c) PSR B1541+09}
	  \fig{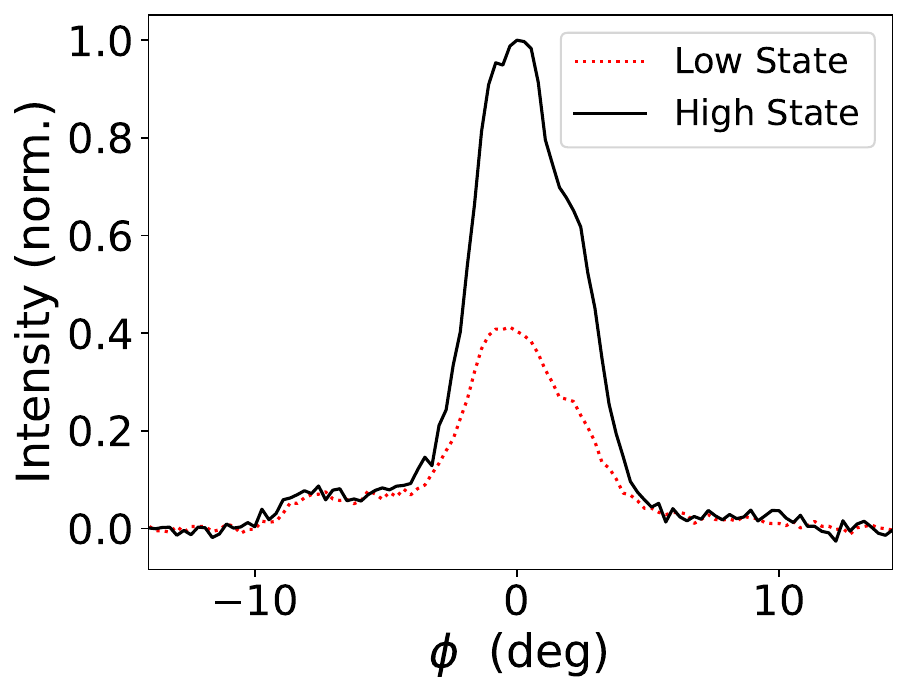}{0.23\textwidth}{(d) PSR B1600$-$49}
         }
\gridline{\fig{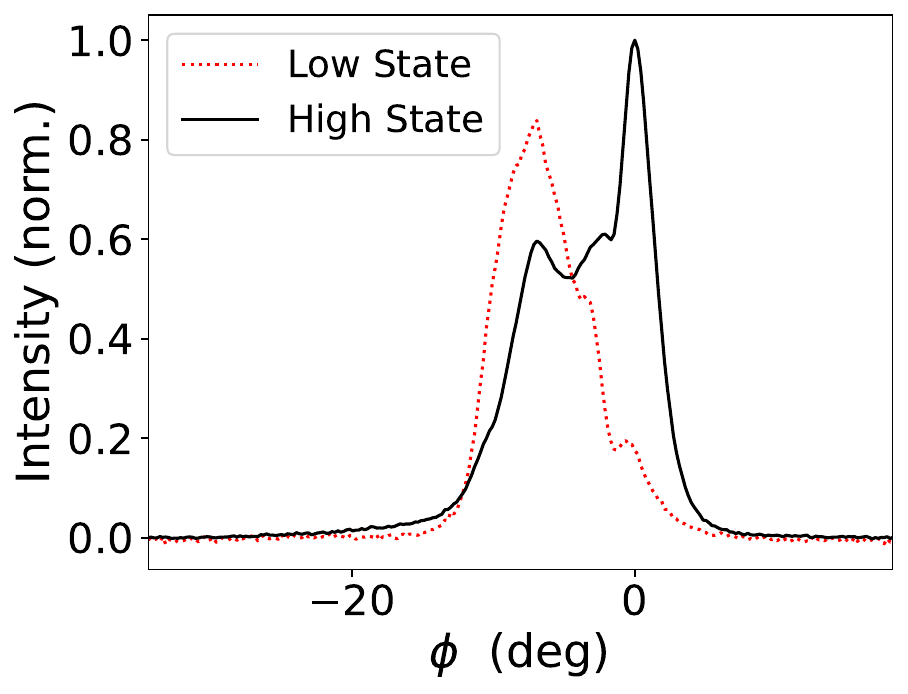}{0.23\textwidth}{(e) PSR B1604$-$00}
	  \fig{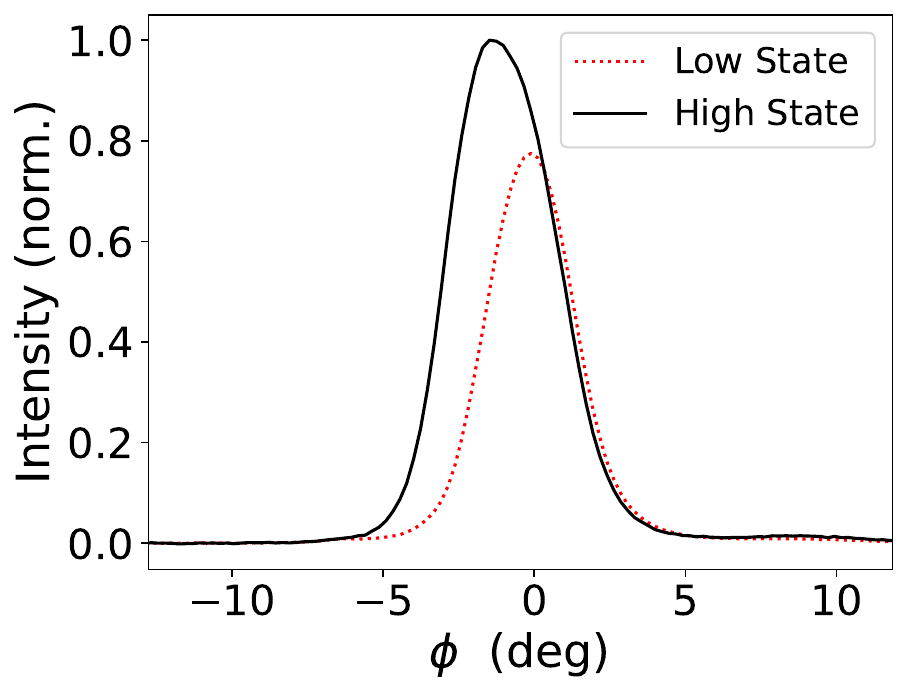}{0.23\textwidth}{(f) PSR B1642$-$03}
	  \fig{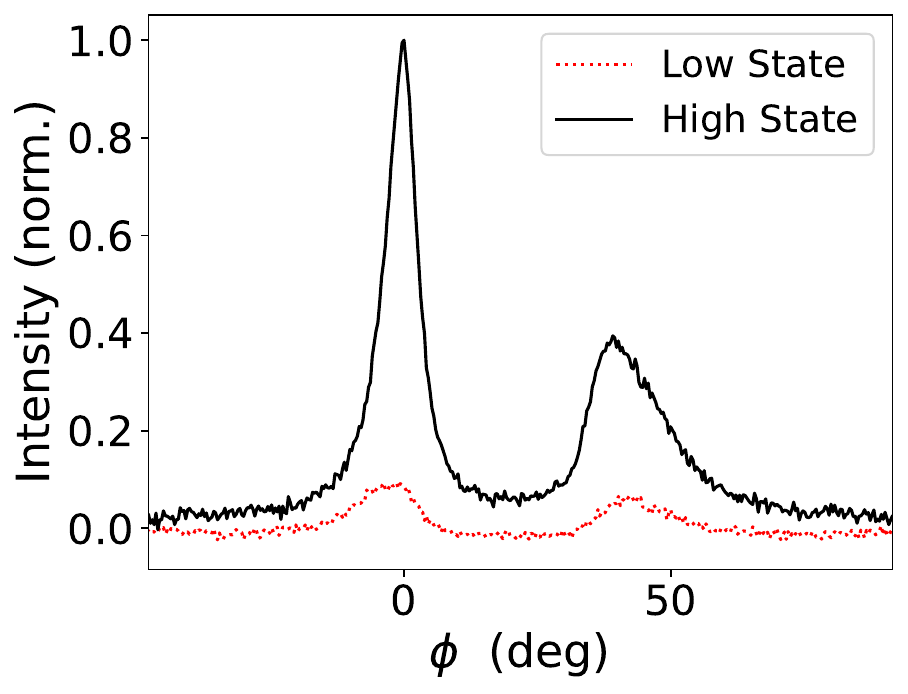}{0.23\textwidth}{(g) PSR B1730$-$37}
         }
\gridline{\fig{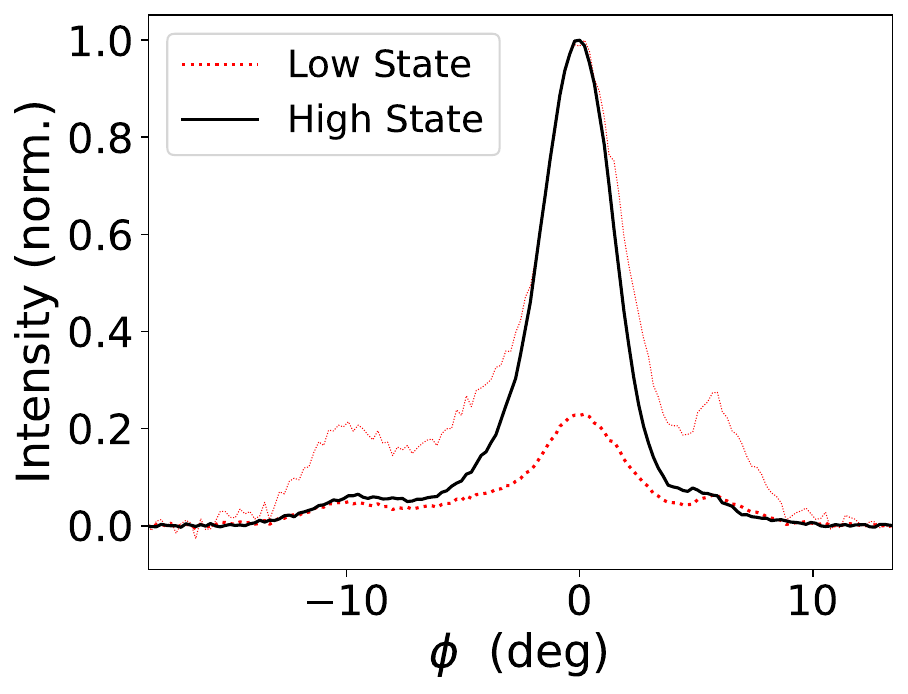}{0.23\textwidth}{(h) PSR B1732$-$07}
	  \fig{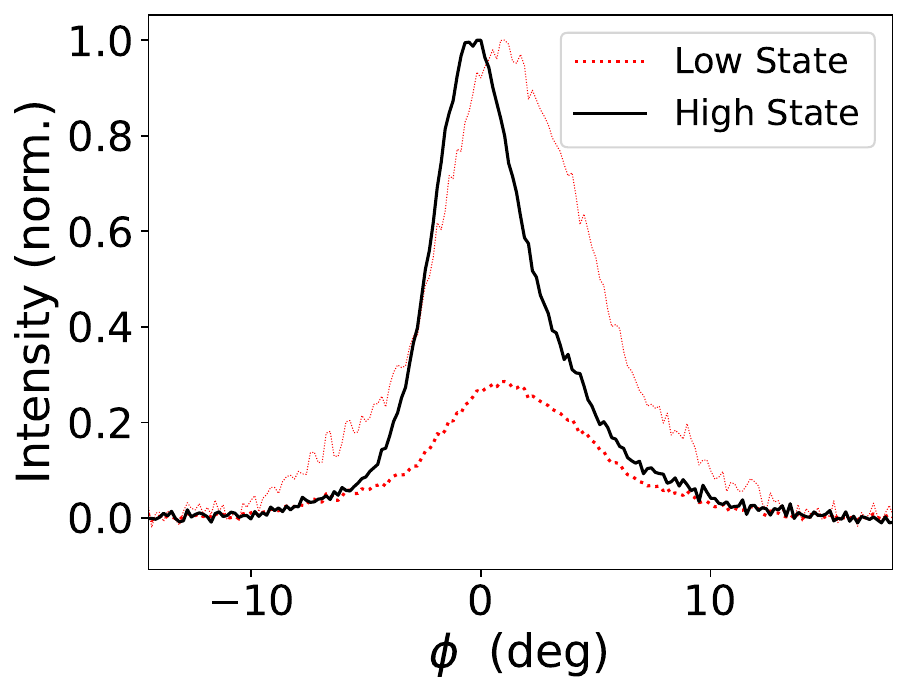}{0.23\textwidth}{(i) PSR B1737$-$39}
	  \fig{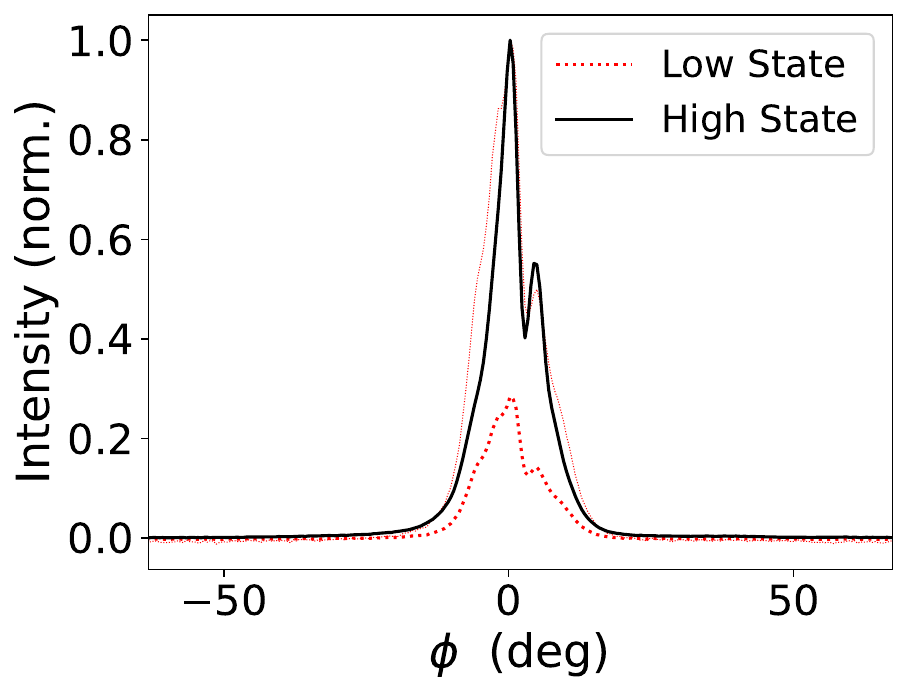}{0.23\textwidth}{(j) PSR B1929+10}
	 }
\gridline{\fig{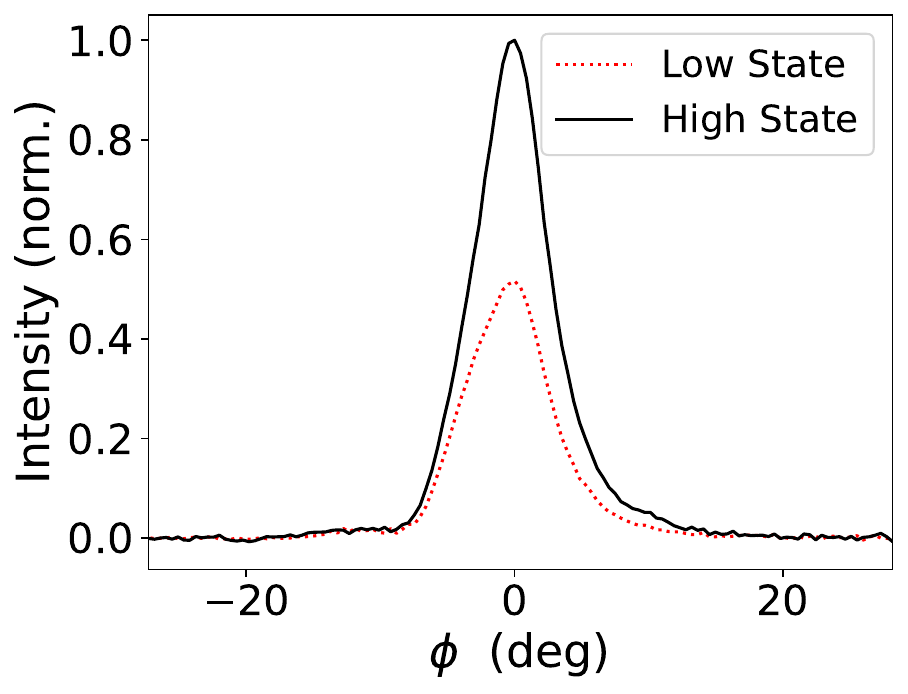}{0.23\textwidth}{(k) PSR B1929+20}
          \fig{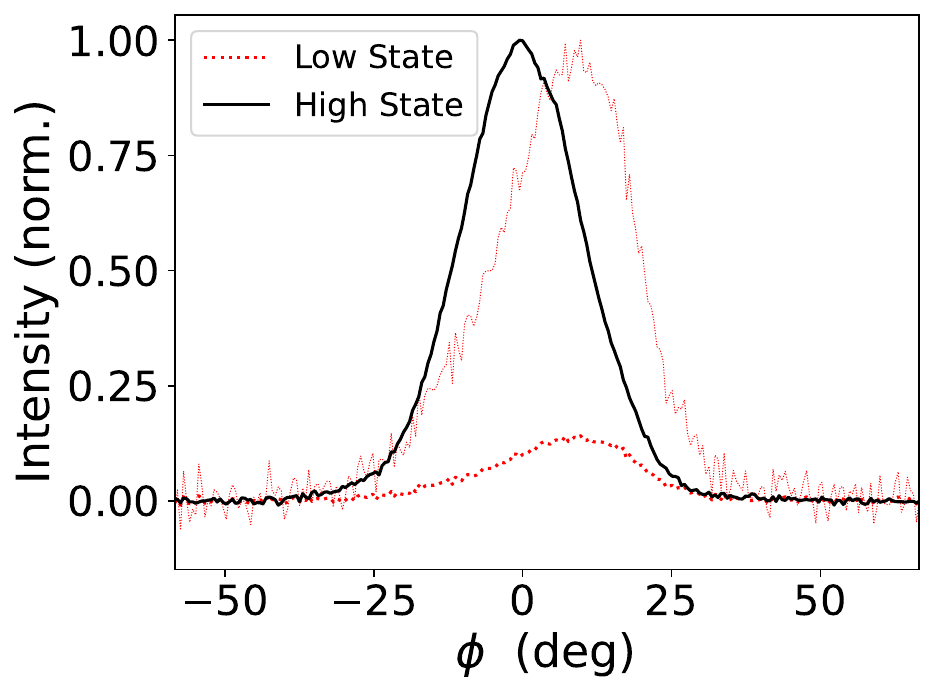}{0.23\textwidth}{(l) PSR B2011+38}
	  \fig{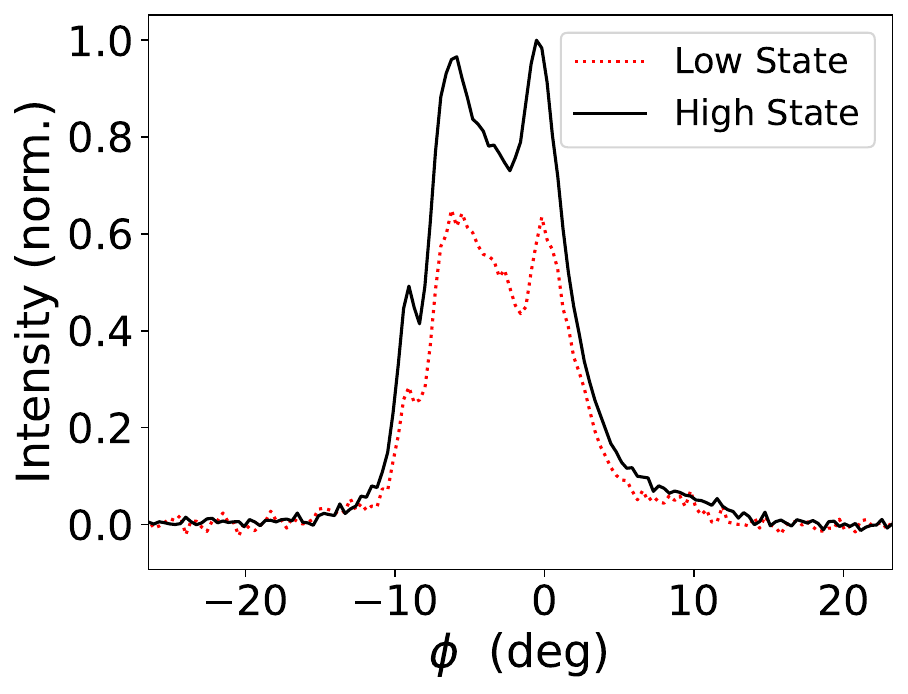}{0.23\textwidth}{(m) PSR B2148+52}
         }
\caption{The average profiles of the brighter (black) and weaker (dotted red)
intensity states of periodic amplitude modulation. In some cases scaled up 
versions of weaker state profiles are also shown to better highlight the 
emission features.
\label{fig:state_prof}}
\end{figure}

\paragraph{\bf B1541+09} This is the least energetic pulsar in our list, with
$\dot{E} < 10^{32}$ erg~s$^{-1}$, and has a relatively wide profile with the
pulsed emission occupying about half the rotation period. The profile consist 
of a central bright core and much weaker conal emission on either side 
\citep{ETS_R93}. Fig.~\ref{fig:state_singl}(e) shows the periodic modulation 
behaviour in the single pulse sequence of the pulsar where the modulation is 
mostly seen in the central core component transitioning regularly between 
bright and weak intensities. The average LRFS shows wide low frequency 
structures below 0.1 cycle/$P$, where the periodic behaviour varies with time 
(see appendix) from more periodic to diffuse features. The phase variations 
associated with the peak frequency shows a relatively flat change across the 
core component which has also been noted by \citet{SWS23}. The cutoff analysis 
shows that the pulsar spends roughly equal amount of times at both states, with
around 56\% of the duration in the brighter state. The average profiles of the 
two states are shown in Fig.~\ref{fig:state_prof}(c), where the core intensity 
in the weaker state is around 30\% level of the brighter state. The weaker 
conal emission remains largely unchanged during the two states. The expanded 
profile of the weaker state shows a bifurcation in the core emission, that has 
also been seen in other core dominated pulsars like PSRs B1933+16 
\citep{MRA16}, B0329+54 \citep{BMR19}, etc. The peak of the weaker profile is 
slightly shifted towards the trailing side, but the overall profile widths of 
the two states remain unchanged (see Table \ref{tab:ampstate}).

\paragraph{\bf B1600$-$49} The pulsar profile is classified as core-cone triple
\citep[T$_{1/2}$,][]{R22}, with a prominent core component and much weaker 
conal emission in the leading side. The single pulse behaviour shows regular 
transitions from a weaker state to bursts of higher intensity state in the core
emission (see Fig.~\ref{fig:state_singl}f). The average LRFS shows a relatively
narrow low frequency feature below 0.05 cycles/$P$ which highlights the 
periodic nature of these transitions. Surprisingly, no such periodic behaviour 
in this pulsar has been reported by \citet{SWS23}, despite high enough 
detection sensitivity. The cutoff analysis shows that the bright states are 
comparatively short lived and seen for a quarter of the observing duration. The
peak intensity of the core emission reduces by 40\% during the weak state while
the conal intensity is similar throughout. The profile width also remains 
unaltered in the two emission states.

\paragraph{\bf B1730$-$37} The pulsar has a wide double (D) profile where the 
two components are clearly separated and connected by low-level bridge 
emission. The single pulses (Fig.~\ref{fig:state_singl}h) are seen as short 
bursts of high intensity emission separated by longer durations of weaker 
emission. The average LRFS shows the presence of a sharp periodic feature with 
a narrow peak, suggesting highly periodic transitions between the two states, 
that differs from the broad periodic features usually associated with this 
phenomenon. The state separation analysis confirms that the bright state is 
less frequent and seen less than 30\% of the observing duration. The average 
profiles of the two states are shown in Fig.~\ref{fig:state_prof}(g) where the 
intensity of the weak state is less than one tenth of the bright state. The two
components are of roughly equal strength in the weak state with negligible 
bridge emission between them. In the bright state the leading component has
more than twice the intensity level of the trailing side with prominent bridge 
emission connecting the two. Much longer emission tails are seen on either side 
of the profile, which is reflected in the much wider profile widths reported in 
Table~\ref{tab:ampstate}. The single pulse emission properties were previously 
studied by \citet{BMM20a}, who using the narrowband (33 MHz) receiver system of
GMRT found the presence of periodic behaviour. However, the detection 
sensitivity was not adequate in this work to detect the low level emission 
from the weak state. The pulsar was also part of the periodic modulation study 
of \citet{SWS23}, where the two dimensional fluctuation spectra 
\citep[2DFS,][]{ES02} has been used to measure the periodic behaviour. In 
addition to measuring the peak modulation frequency, the 2DFS also estimates 
the longitudinal separation ($P_2$) between adjacent drift bands, based on the 
phase variations seen in the LRFS. This study reports the presence of non-drift
periodic modulation in the trailing component of this pulsar but drift 
behaviour in the leading component with $P_2$ = 119$\degr$. The single pulse 
sequence in Fig.~\ref{fig:state_singl}(h) do not show any evidence of drifting 
in the leading component, while $P_2$ being much larger than the component 
width likely precludes the presence of clear drift bands. We conclude that 
despite the highly periodic nature of the single pulse modulation there is no 
clear evidence of subpulse drifting in this pulsar and both components exhibit 
periodic amplitude modulation.

\paragraph{\bf B1732$-$07} The pulsar has a triple (T) profile with a strong 
central core component and weaker conal emission on both sides. The modulation 
behaviour is clearly seen in the single pulse sequence of 
Fig.~\ref{fig:state_singl}(i) where the central component regularly transitions
between the weak and bright states. The average LRFS shows wide low-frequency 
feature below 0.1 cycles/$P$ highlighting the periodic nature of the intensity 
fluctuations. The pulsar spends roughly equal amount of times in the two 
states. The average profiles of the two emission states (see 
Fig.~\ref{fig:state_prof}h) show the core emission in the weaker state to be 
wider with peak intensity around four times lower than the bright profile. The 
conal emission as well as the profile widths (Table~\ref{tab:ampstate}) remain 
unchanged in the two states. The modulation properties of this pulsar 
have been reported in \citet{SWS23}, who suggest the emission behaviour to be 
subpulse drifting with $P_2$ = 16$\degr$. However, the single pulse sequence 
does not show any drift bands and $P_2$ is larger than the core width, which 
makes it much more likely for the periodic behaviour to be periodic amplitude 
modulation.

\paragraph{\bf B1929+10} The pulsar profile exhibits a merged double component 
structure along with an interpulse emission and low-level bridge emission 
connecting the main pulse with the interpulse \citep{RR97}. The periodic 
amplitude modulation behaviour in this source has been investigated by 
\citet{KYP21} using sensitive observations with FAST. The study found the
modulation features in the main pulse and more prominently in the interpulse
with identical periodicities and a constant phase relationship between them. 
The phase variations corresponding to the peak frequency of fluctuation 
remained flat across the interpulse but showed variations across the main 
pulse. In the interpulse region the intensities of the two emission states had
clear statistical boundary and they were separated to form profiles that showed 
similar polarization behaviour. Our observations were not sensitive enough to 
detect the single pulse emission from the interpulse region thereby limiting 
the state separation analysis to the main pulse. Fig.~\ref{fig:state_singl}(l) 
shows a short sequence of the single pulses with bright and weak emission 
states. The bright state is more dominant in the main pulse lasting around 70\%
of the duration. The ratio between the intensities of the weak state and 
the bright state is around 0.3 and their emission windows coincide. The phase 
variations obtained from the frequency peak in the LRFS remain flat across the 
main pulse (see appendix) which differs from the higher frequency FAST 
measurements.

\paragraph{\bf B1929+20} The pulsar has a single component profile and the 
periodic modulation in the single pulse sequence (Fig.~\ref{fig:state_singl}m)
is being reported for the first time. The average LRFS shows a typical wide 
low-frequency feature with peak frequency located below 0.1 cycles/$P$ and 
relatively flat phase variations across the emission window (see appendix). The
profiles of the two states are similar (Fig.~\ref{fig:state_prof}k) with the 
brighter state having around twice the intensity of the weak state, and it is 
seen for 36\% of the observing duration. 

\paragraph{\bf B2148+52} The average profile in this pulsar has a complex shape
with two merged peaks, a slight projection at the leading side and an elongated
trailing tail. The periodic modulation behaviour is seen in the single pulse 
sequence (Fig.~\ref{fig:state_singl}) as regular transitions between the bright
and the weak states. The average LRFS shows the presence of a relatively 
narrow peak around 0.1 cycles /$P$ with periodicity $P_M$ = 8.8$\pm$0.5 $P$, 
that constitute a new detection of periodic amplitude modulation. The cutoff 
analysis manages to separate the two emission states and the low frequency 
feature is clearly seen in the average FFT of the on-off sequence. The bright 
state appears for longer durations with 70\% relative abundance. The average 
profiles of the two states are similar with the weaker state having 60-70\% 
intensity of that in the bright state.

\subsection{Periodic Modulations with Phase Shift} \label{sec:Phs_Sft}

\centerwidetable
\begin{deluxetable}{ccccccccccc}
\tablecaption{Emission States of Periodic Amplitude Modulation
\label{tab:ampstate}}
\tabletypesize{\footnotesize}
\tablewidth{0pt}
\tablehead{
 \colhead{Name} & \colhead{Type} & \colhead{Cutoff} & \multicolumn{3}{c}{\underline{Bright Emission State}} & \multicolumn{3}{c}{\underline{Weak Emission State}} & \multicolumn{2}{c}{\underline{Intensity Ratio (S$_{\rm W}$/S$_{\rm B}$)}} \\
   &   &   & \colhead{Abundance} & \colhead{$W_{5\sigma}$} & \colhead{$W_{10}$} & \colhead{Abundance} & \colhead{$W_{5\sigma}$} & \colhead{$W_{10}$} & \colhead{Peak} & \colhead{Total} \\
   &   & \colhead{($\cal{F}_{\rm B}$)} & \colhead{(\%)} & \colhead{(deg)} & \colhead{(deg)} & \colhead{(\%)} & \colhead{(deg)} & \colhead{(deg)} &  &  }
\startdata
  B0450+55  & Phase Mod & 0.38 & 37.0$\pm$1.2 & 51.9$\pm$1.1 & 31.4$\pm$1.1 & 63.0$\pm$1.5 & 37.7$\pm$1.1 & 29.8$\pm$1.1 & 0.733$\pm$0.005 & 0.540$\pm$0.003 \\
 B0905$-$51 & Phase Mod & -0.46 & 68.7$\pm$1.4 & 141.0$\pm$0.9 & 99.2$\pm$0.9 & 31.3$\pm$0.9 & 94.7$\pm$0.9 & --- & 0.502$\pm$0.003 & 0.699$\pm$0.004 \\
  B1541+09  & Phase Stat & -0.16 & 56.4$\pm$1.4 & 160.4$\pm$0.5 & --- & 43.6$\pm$1.2 & 162.2$\pm$0.5 & --- & 0.286$\pm$0.005 & 0.376$\pm$0.001 \\
 B1600$-$49 & Phase Stat & 0.92 & 24.4$\pm$1.1 & 17.9$\pm$0.5 & --- & 75.6$\pm$1.9 & 16.3$\pm$0.5 & --- & 0.412$\pm$0.009 & 0.47$\pm$0.09 \\
 B1604$-$00 & Phase Mod & -0.84 & 84.0$\pm$1.5 & 35.1$\pm$0.4 & 16.7$\pm$0.4 & 16.0$\pm$0.7 & 18.5$\pm$0.4 & 16.4$\pm$0.4 & 0.84$\pm$0.005 & 0.804$\pm$0.004 \\
 B1642$-$03 & Phase Mod & 1.20 & 19.4$\pm$0.9 & 20.7$\pm$0.5 & 7.0$\pm$0.5 & 80.6$\pm$1.9 & 22.9$\pm$0.5 & 6.3$\pm$0.5 & 0.775$\pm$0.001 & 0.65$\pm$0.03 \\
 B1730$-$37  & Phase Stat & 0.68 & 28.7$\pm$1.1 & 141.4$\pm$0.7 & 107.6$\pm$0.7 & 71.3$\pm$1.7 & 61.5$\pm$0.7 & --- & 0.094$\pm$0.006 & 0.074$\pm$0.001 \\
 B1732$-$07 & Phase Stat & -0.12 & 54.6$\pm$1.6 & 21.7$\pm$0.4 & --- & 45.4$\pm$1.5 & 21.6$\pm$0.4 & --- & 0.230$\pm$0.003 & 0.34$\pm$0.03 \\
 B1737$-$39 & Phase Mod & 0.98 & 26.4$\pm$1.1 & 19.2$\pm$0.3 & 12.3$\pm$0.3 & 73.6$\pm$1.9 & 20.7$\pm$0.3 & 17.6$\pm$0.3 & 0.285$\pm$0.006 & 0.42$\pm$0.04 \\
  B1929+10  & Phase Stat & -0.46 & 68.2$\pm$1.6 & 200.0$\pm$1.0 & 22.5$\pm$1.0 & 31.8$\pm$1.0 & 184.1$\pm$1.0 & 22.3$\pm$1.0 & 0.284$\pm$0.001 & 0.354$\pm$0.003 \\
  B1929+20  & Phase Stat & 0.36 & 36.6$\pm$1.2 & 27.6$\pm$0.9 & 13.7$\pm$0.9 & 63.4$\pm$1.6 & 26.3$\pm$0.9 & 14.0$\pm$0.9 & 0.516$\pm$0.004 & 0.559$\pm$0.004 \\
  B2011+38  & Phase Mod & 0.36 & 40.6$\pm$1.4 & 64.3$\pm$1.0 & 43.2$\pm$1.0 & 59.4$\pm$1.7 & 54.1$\pm$1.0 & --- & 0.141$\pm$0.003 & 0.174$\pm$0.001 \\
  B2148+52  & Phase Stat & -0.56 & 71.8$\pm$2.7 & 26.3$\pm$0.7 & 18.7$\pm$0.7 &  28.2$\pm$1.7 & 23.0$\pm$0.7 & --- & 0.648$\pm$0.014 & 0.677$\pm$0.008 \\
\tableline
\enddata
\end{deluxetable}

\paragraph{\bf B0450+55} The pulsar has a triple (T) profile with three merged
components, asymmetrically extending towards the trailing side. The periodic 
modulation is primarily seen in the trailing component showing regular 
transitions from the bright to the weak state. The average LRFS shows a wide 
low-frequency feature peaking around 0.1 cycles/$P$, and there are indications 
of phase variations with positive slope associated with the periodic 
fluctuations (see appendix). The pulse sequence shows the effects of 
interstellar scintillation with fluctuations in the intensity levels, which 
made it challenging for the statistical state separation analysis. The average 
FFT of the on-off sequence shows the presence of the low frequency periodic 
feature and 37\% of the single pulses comprises of the bright state. The phase 
variation in the periodic modulation behaviour is also evident in the the 
bright and weak state profiles (Fig.~\ref{fig:state_prof}a). The trailing conal
component has lower intensity and is clearly separated from the core in the 
weak state profile. On the contrary the trailing side intensity increases and 
merges with the core in the bright state profile. The peak intensity of the 
weak state profile is around 75\% of the bright state profile, while in the 
trailing side the emission intensity is more than three times higher in the 
bright state.

\paragraph{\bf B0905$-$51} The pulsar has three merged components in a 
relatively wide average profile ($W_{5\sigma}\sim$136$\degr$), where the 
trailing component is the most dominant feature with more than twice the 
intensity level of the central core component and five times higher than the 
leading cone. The modulation behaviour is seen in the trailing side with the 
emission window becoming wider during the bright state (see 
Fig.~\ref{fig:state_singl}d). The average fluctuation spectrum shows a somewhat
diffuse low frequency feature with peak frequency below 0.05 cycles/$P$, which 
is typical of the periodic amplitude modulation behaviour. The average profiles
of the two emission states also highlight the modulation behaviour with the 
bright state profile being wider on the trailing side and its maximum intensity
level being two times higher than the weak state profile peak. The central 
component has around 15\% lower intensity in the weaker state but the leading 
component is identical in the two profiles. The brighter state is more 
prevalent in the single pulse sequence and seen around 70\% of the duration of 
our observations. The modulation properties of this pulsar have also been 
measured by \citet{SWS23}, who suggested the behaviour to represent subpulse 
drifting due to non-zero phase variations resulting in $P_2$ = 42$\degr$. 
However, our analysis shows that the estimated phase variations are not a 
result of systematic drift motion of the subpulses but rather due to the 
widening and narrowing of the pulse window in the two emission states of 
periodic amplitude modulation.

\paragraph{\bf B1604$-$00} The three components in the pulsar profile are 
merged together with the central core emission having lower intensity than the
two conal components on either side and serving as a bridge between them. The 
single pulse sequence of Fig.~\ref{fig:state_singl}(g) shows a phase 
misalignment between the leading and trailing sides, i.e. when the bright 
emission is seen in the leading side the trailing side has weaker emission and 
vice versa, similar to the emission behaviour of PSR B1946+35 \citep{MR17,
CWK25}. The average LRFS has a zero frequency feature in addition to the peak 
frequency around 0.03 cycles/$P$ due to periodic amplitude modulation. The 
phase variations associated with the peak modulation frequency in the LRFS also
show a 180$\degr$ jump between the leading and trailing components (see 
appendix). The bright state corresponds to the case when the trailing side has 
higher intensity and lasts for longer durations, seen around 85\% of observing 
time. In the shorter lived weak state the leading component is dominant. The
leading component of the weak state profile is five times brighter than its 
trailing side and about 84\% level of the trailing component in the bright 
state profile. \citet{SWS23} classified the periodic modulation feature in this
source as non-drifting behaviour with no measurable $P_2$.

\paragraph{\bf B1642$-$03} The pulsar with a single component profile shows two 
distinct emission states in the single pulse sequence with the emission window
becoming wider at the leading side during the bright state (see 
Fig.~\ref{fig:B1642_singl}, left panel). The average fluctuation spectrum in 
Fig.~\ref{fig:B1642_permod}, left panel, shows the presence of a wide double 
peaked feature below 0.1 cycles/$P$ due to periodic transitions between the two 
states. The change in the emission window during periodic amplitude modulation
is also seen in the LRFS (see appendix) where the maximum modulation occurs in 
the leading side ahead of the profile peak and in certain sequences, when the 
peak frequency has higher SNR, there are non-zero phase variations with a 
negative slope. The bright state is short-lived and seen in 20\% of the 
observed pulses. The weak state profile is shifted to the trailing side of the 
bright state profile with maximum intensity around 77.5\% of that in the bright
state peak. \citet{SWS23} reported the presence of subpulse drifting in this 
pulsar with $P_2 = 49\degr$ which is much larger than the profile widths. Our 
analysis show that there is no clear evidence of subpulse drifting and the 
estimated $P_2$ is due to the different emission windows of the two states 
associated with periodic amplitude modulation.

\paragraph{\bf B1737$-$39} A single component profile is seen in this pulsar
with a slight scattering tail near the trailing side at the 550-750 MHz 
frequency band. The single pulse sequence shows clear bursts of high intensity
emission interspersed between the lower intensity states (see 
Fig.~\ref{fig:state_singl}j). The average fluctuation spectra shows two 
periodic features, a relatively sharp feature with full width at half maximum 
(FWHM) $\sim$ 0.02 cycles/$P$ at low frequencies below 0.05 cycles/$P$, and a 
relatively wider structure with FWHM $\sim$ 0.08 cycles/$P$ centered around the 
higher frequency of 0.1 cycles/$P$. The state separation analysis show that the
average FFT from the on-off sequence reproduces both these periodic features, 
suggesting that two different types of periodicity are associated with periodic
amplitude modulation in this pulsar. The bright states are relatively short 
lived and is seen around 26\% of the observing duration. The bright pulses are 
shifted towards the leading side of the emission window. The weak state profile
is wider and extended towards the trailing side and has peak intensity less 
than 30\% of that in the bright profile. The pulsar was also analysed by 
\citet{SWS23}, but no periodic behaviour was reported in this work. 

\paragraph{\bf B2011+38} The average profile of this pulsar has a single 
component and the pulse sequence in Fig~\ref{fig:state_singl}(o) shows large 
fluctuations of intensities in a fairly haphazard manner without any clear 
visible patterns. A prominent wide low frequency feature centered between 
0.03-0.04 cycles/$P$ is seen in the average fluctuation spectrum, revealing the
presence of periodic amplitude modulation. The cutoff analysis was able to 
separate the two emission states showing the bright state profile with peak 
intensity around 7 times higher than the weak state profile peak. The single 
pulse distribution is more even between the two states with the pulsar spending
around 40\% of the observing duration in the bright state and the remaining 
60\% duration shows weak emission. A scaled up version of the weak state 
profile in Fig.~\ref{fig:state_prof}(l) reveals a phase shift between the two 
profiles with the weak state profile peak shifted by around 10$\degr$ longitude
towards the trailing side with respect to the bright state profile. The 
periodic behaviour from this source has also been reported by \citet{WES06} who
suggested the modulation to represent longitude stationary drift behaviour.

\subsection{Intermittent Periodic Modulation} \label{sec:Int_Var}

\paragraph{\bf B0136+57} The pulsar has a single component profile where the 
single pulses in Fig.~\ref{fig:state_singl}(a) show regular transitions between
bright pulses and weak emission. The time average fluctuation spectrum shows a 
diffuse wide feature between 0.1-0.2 cycles/$P$ which is prominent near the 
later parts of the observations (see Fig.~\ref{fig:lrfs_int} in appendix), and 
constitute a new detection of periodic amplitude modulation from this pulsar. 
The LRFS from a short sequence shows a wider peak exhibiting non-zero phase 
variations with positive slope near the leading side of the profile.

\paragraph{\bf B0450$-$18} A triple profile is seen in this pulsar with three
distinct components \citep{ETS_R93}. The single pulse sequence is shown in 
Fig.~\ref{fig:state_singl}(b) where regular bursts of higher intensity emission
are seen primarily in the leading and trailing components. The average LRFS
shows the presence of a weak diffuse feature suggesting that the intensity 
modulations in this source do not exhibit regular periodic behaviour. The 
pulsar was observed by \citet{SWS23} who do not report the presence of any 
periodic modulation.

\paragraph{\bf B1917+00} The pulsar also has a triple profile \citep{ETS_R93}
consisting of a dominant central core and prominent conal regions on either 
side merged with the core component. The single pulses (see 
Fig.~\ref{fig:state_singl}k) have short bursts of high intensity emission 
primarily in the central core interspersed between longer durations of weak 
emission state. The time evolution of the average LRFS shows that the high 
intensity states do not appear in a periodic manner apart from a short sequence
of a few hundred pulses where a narrow periodic feature is seen near 0.1 
cycles/$P$. No periodic behaviour was reported from this source by 
\citet{SWS23}.

\paragraph{\bf B2334+61} The pulsar has a composite profile shape without any
clear distinction between components. The single pulses shown in 
Fig.~\ref{fig:state_singl}(p) have several instances of high-intensity emission
scattered throughout the pulse window without any discernible pattern, similar 
to the emission from PSR B2011+38. The average fluctuation spectrum shows a 
relatively weak peak around frequencies of 0.05 cycles/$P$, that has not been 
reported previously. The periodicity of the fluctuations becomes prominent at 
several short intervals and the corresponding LRFS shows relatively narrow 
peaks (see Fig.~\ref{fig:lrfs_int} in appendix), where the periodic behaviour 
is limited to the trailing side of the profile (red dots in the bottom window 
of LRFS showing the amplitude of spectral feature with significant detection).

\section{Discussion} \label{sec:Discuss}

The single pulse behaviours of periodic amplitude modulation despite showing 
considerable variations in different sources have certain distinguishing 
features. The periodicity results from transitions between two emission states 
with different intensity levels. In certain cases the emission windows of the 
two states can have different widths or the intensity changes only at specific 
longitudinal locations, and this can cause non-zero phase variations in the 
LRFS. But none of these effects are associated with any systematic shift of 
subpulses across the emission window, which marks the major observational 
difference between periodic amplitude modulation and subpulse drifting. The 
statistical scheme for separating the two emission states of periodic amplitude
modulation gives a good demonstration of this effect. In almost all cases, 
apart from PSR B0450+55 whose single pulse energy distribution was affected by 
interstellar scintillation, where significant detection of the periodic feature
were recorded, we were able to reproduce the periodic feature from the FFT of 
the `0/1' sequence corresponding to the two states with the same sensitivity as
the average LRFS. In this sequence all information about subpulse structure 
within the single pulses are washed away demonstrating that the periodicity 
arises purely from the transition between the two states. 

Several differences have emerged among the physical properties of subpulse
drifting, periodic amplitude modulation and periodic nulling, since their 
identification as separate phenomena \citep[see][for a discussion]{MBM24a}. The
major differences are related to the energetics, where drifting is limited to 
pulsars with $\dot{E} < 5\times10^{32}$ ergs~s$^{-1}$, while the periodic 
amplitude modulation and periodic nulling are seen over a wider $\dot{E}$ 
range, and particularly pulsars with higher $\dot{E}$ are associated with 
periodic amplitude modulation \citep{BMM17,BMM20a}. The drifting periodicities 
also appear to be inversely correlated with $\dot{E}$, despite aliasing effect 
making it difficult to estimate the actual periodicities in certain instances 
\citep{BMM16,BMM19}. The anti-correlation between the two quantities emerges
from the partially screened gap nature of the inner acceleration region where
the outflowing plasma is generated and undergoes drift, and the drifting 
periodicity $P_3\propto\dot{E}^{-0.5}$ \citep{GMG03,BMM16}. The other 
difference is related to the localization of these two phenomena within the 
pulsed emission window. The periodic amplitude modulation and periodic nulling
are seen in all emission components, even though in some cases only a part of 
the window participates in periodic amplitude modulation. The subpulse drifting
on the other hand is only seen in the conal components and the central core 
component which can be associated with the central sparking region in the polar
cap do not drift \citep{BMM22b,BMM23a}.

\begin{figure}
\epsscale{1.1}
\plottwo{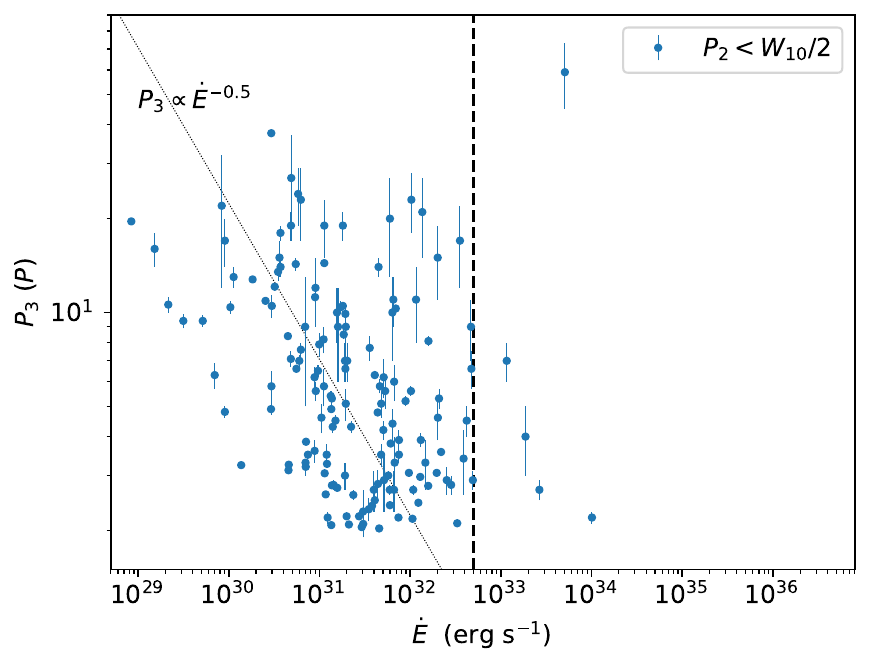}{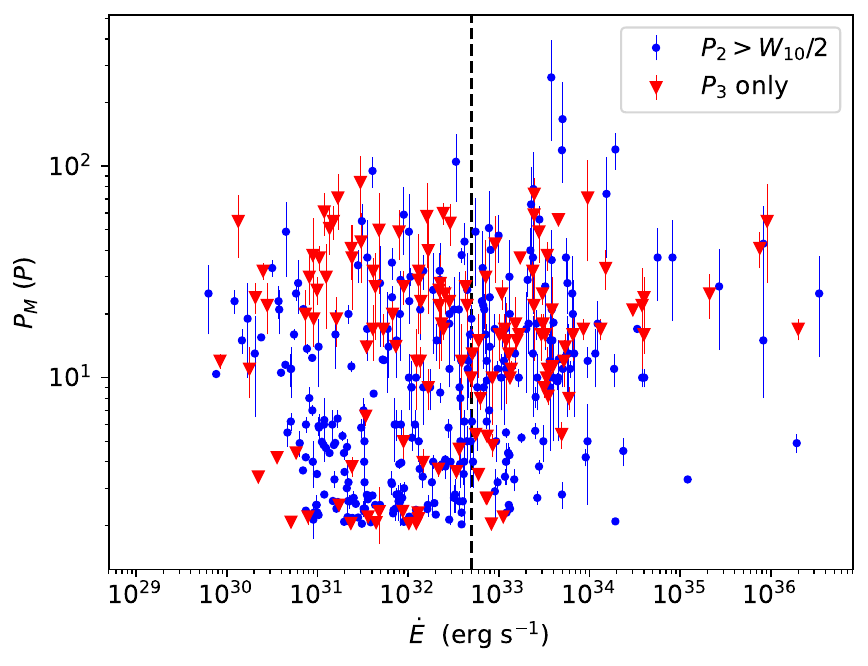}
\caption{The distribution of the different categories of periodic behaviour 
with $\dot{E}$, from the survey of pulsars reported in \citet{SWS23}. The 
left panel shows the pulsars with $P_2 < W_{10}/2$ which likely represent the 
subpulse drifting population along with the line showing the expected 
dependence, $P_3\propto\dot{E}^{-0.5}$. The right panel shows pulsars with 
$P_2 > W_{10}/2$ (green) representing periodic modulations with phase shift, 
and $P_3$-only behaviour (red) belonging to phase stationary behaviour and 
periodic nulling. The dashed vertical line in both plots correspond to $\dot{E}
= 5\times10^{32}$~erg~s$^{-1}$ considered as the upper limit for drifting 
behaviour in previous studies.
\label{fig:Song_mod}}
\end{figure}

The number of pulsars showing periodic behaviour has undergone a significant 
increase recently due to the extensive work of \citet{SWS23} who studied 
pulsars observed in the Thousand-Pulsar-Array programme on MeerKAT and reported
more than five hundred sources showing some form of periodicity in the single 
pulse sequence. The study used the 2DFS technique to measure the periodicity 
and distinguished the periodic amplitude modulation and periodic nulling from 
subpulse drifting by using the estimated $P_2$ value. The periodic behaviour 
with  phase variations across the emission window, i.e. non-zero $P_2$, has 
been classified as subpulse drifting, while the remaining cases were classified
as $P_3$-only behaviour. Based on this classification the the study found that 
in the lower energetic pulsars the anti-correlation between the drifting 
periodicity and pulsar energetics was still relevant, but a large number of 
high energetic sources with $\dot{E} > 5\times10^{32}$~erg~s$^{-1}$ also showed
drifting behaviour, blurring the physical differences between these three 
phenomena. 

In this work we have shown that there exist periodic modulations with phase 
shifts across the emission window, and likely non-zero $P_2$ values in the 
2DFS, without any signatures of subpulse drifting in the single pulse sequence.
Indeed the periodic behaviour of several sources studied here have been 
classified as subpulse drifting in the earlier work. So, it is clear that using
measurements from the fluctuation spectra alone are not sufficient in 
categorising the nature of the periodic behaviour, and careful inspection of 
the single pulse sequence is necessary, which may not be always possible when
dealing with large surveys. Nonetheless, we make an initial attempt to classify
the periodic behaviour from this large sample based on the nature of the phase 
variations. In the case of subpulse drifting, $P_2$ represents the average 
separation between adjacent drift bands, and for any systematic drifting 
behaviour the $P_2$ can be clearly measured when at least two adjacent drift 
bands are visible within the emission window, i.e $P_2 < W/2$. A large value of
$P_2$, more than $W/2$, is often related to the different window sizes of the 
two states of periodic amplitude modulation. 

We have compared the $P_2$ values of modulation from \citet{SWS23} with the 
$W_{10}$ measurements of these pulsars reported in \citet{PKJ21}, and divided 
the periodic behaviour into three groups, the first with $P_2 < W_{10}/2$ which
is likely the drifting population, the second with $P_2 > W_{10}/2$ which 
represents the periodic modulations with phase shifts, and the final group with 
no $P_2$ measurement representing pulsars with phase stationary modulation and 
periodic nulling. In less than 5 pulsars $W_{10}$ measurements were not 
available, while in many cases more than one type of periodic behaviour were 
reported and we classified them separately. We found 152 instances of $P_2 <
W_{10}/2$, 295 cases of $P_2 > W_{10}/2$ and 134 $P_3$-only periodic features. 
Fig.~\ref{fig:Song_mod} shows the distribution of the modulation periodicities 
of these three groups with $\dot{E}$ and seem to follow the established trend. 
The drifting pulsars with $P_2 < W_{10}/2$ (left panel of 
Fig.~\ref{fig:Song_mod}) are primarily seen in pulsars with $\dot{E} < 
5\times10^{32}$ erg~s$^{-1}$. The anti-correlation between the drifting 
periodicity and $\dot{E}$ also emerges from this sample. The distribution of 
the other two groups (right panel in Fig.~\ref{fig:Song_mod}) do not show any 
clear dependence on $\dot{E}$. We emphasize that these are tentative results 
and proper classification would require more detailed study of the single pulse
sequence of each source. For example, five pulsars J1056$-$6258, J1114$-$6100, 
J1350$-$5115, J1524$-$5706 and J1918+1444, with high $\dot{E}$ also have $P_2 <
W_{10}/2$ and further studies are required to check if they are indeed 
exceptions or show special types of phase shifted periodic modulations. A group
of pulsars in the lower left side of the right panel shows the anti-correlated 
periodic behaviour with $\dot{E}$ that may belong to subpulse drifting 
category, with specific line of sight traverse across the sparking pattern.

\begin{acknowledgments}
DM acknowledges the support of the Department of Atomic Energy, Government of
India, under project no. 12-R\&D-TFR-5.02-0700. This work was supported by the
grant 2020/37/B/ST9/02215 of the National Science Centre, Poland.
\end{acknowledgments}

\bibliography{reflist}{}
\bibliographystyle{aasjournalv7}

\begin{figure}
\gridline{\fig{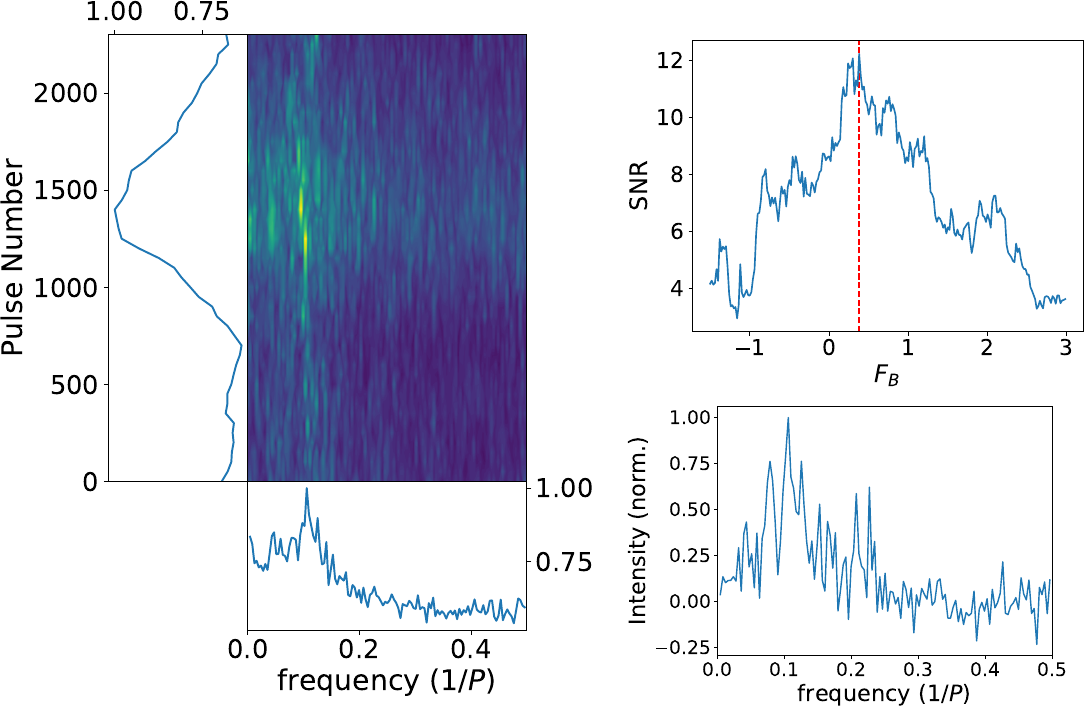}{0.46\textwidth}{PSR B0450+55}
          \fig{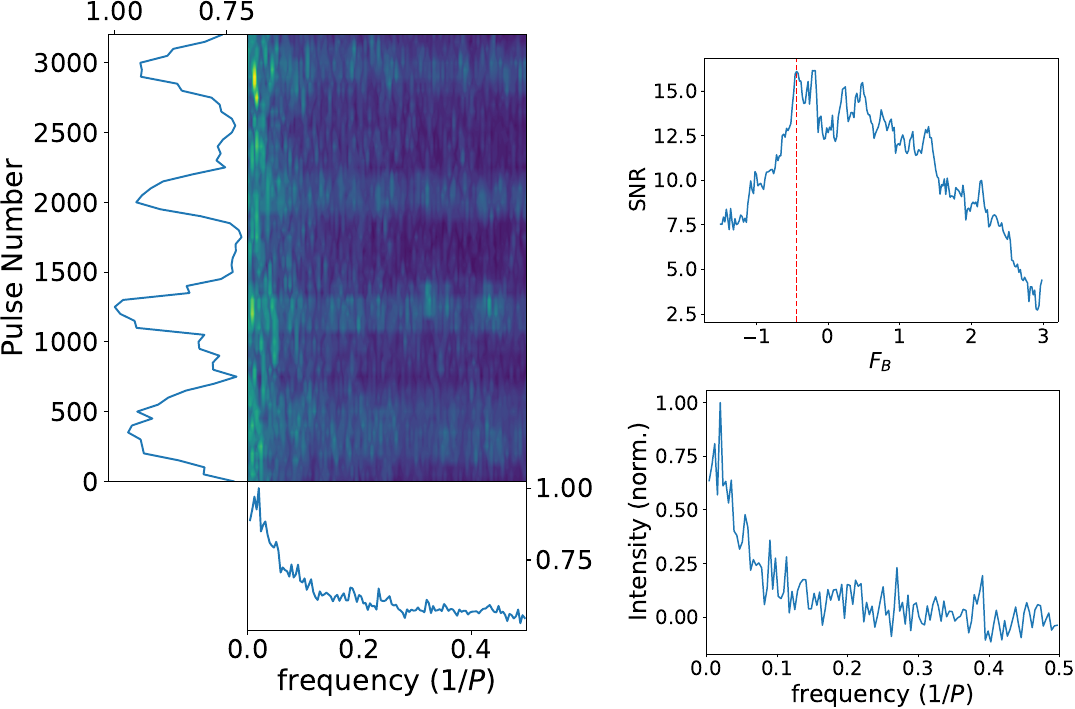}{0.46\textwidth}{PSR B0905$-$51}
         }
\gridline{\fig{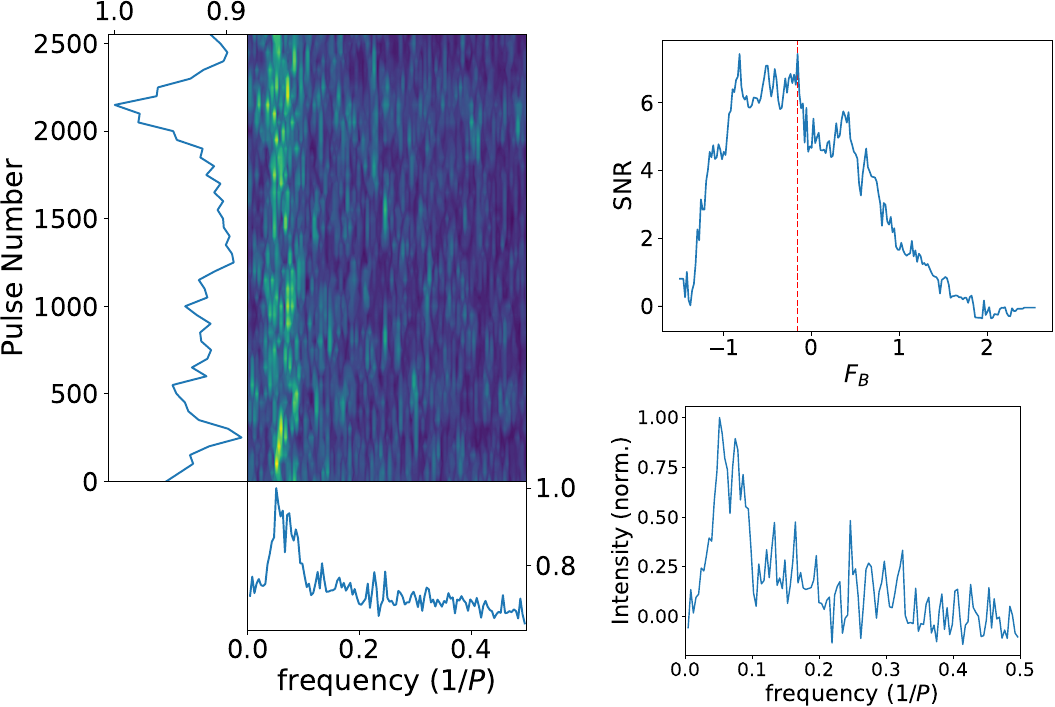}{0.46\textwidth}{PSR B1541+09}
          \fig{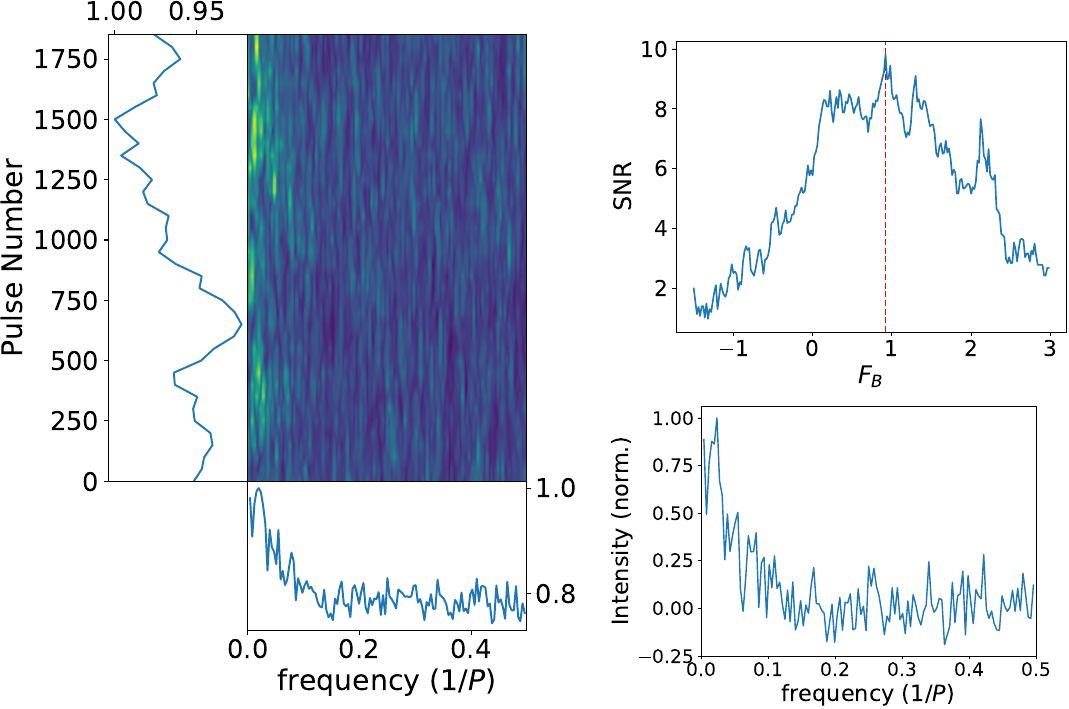}{0.46\textwidth}{PSR B1600$-$49}
         }
\gridline{\fig{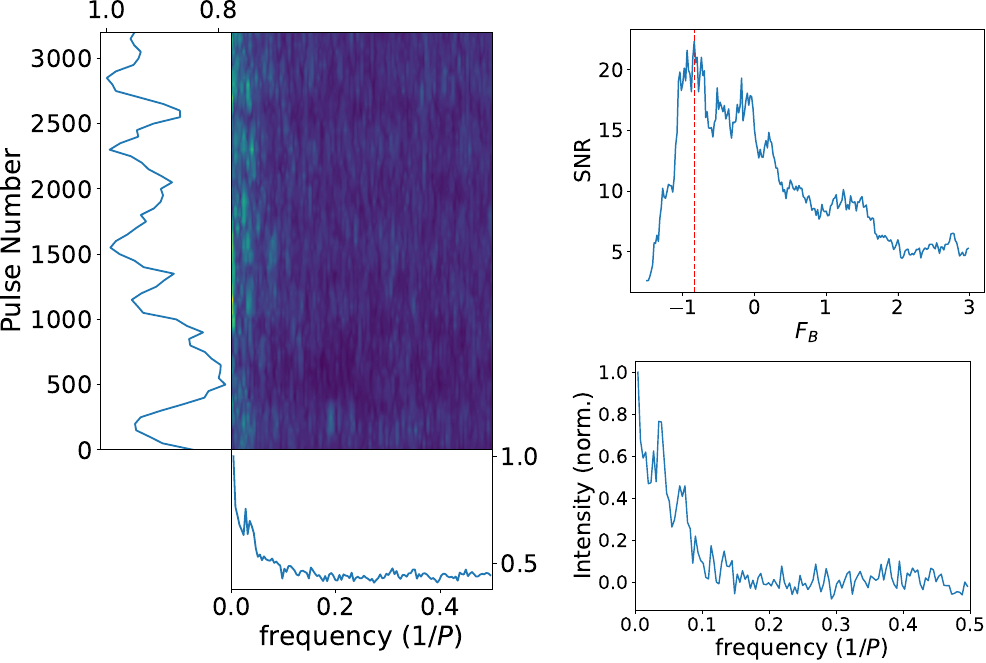}{0.46\textwidth}{PSR B1604$-$00}
	  \fig{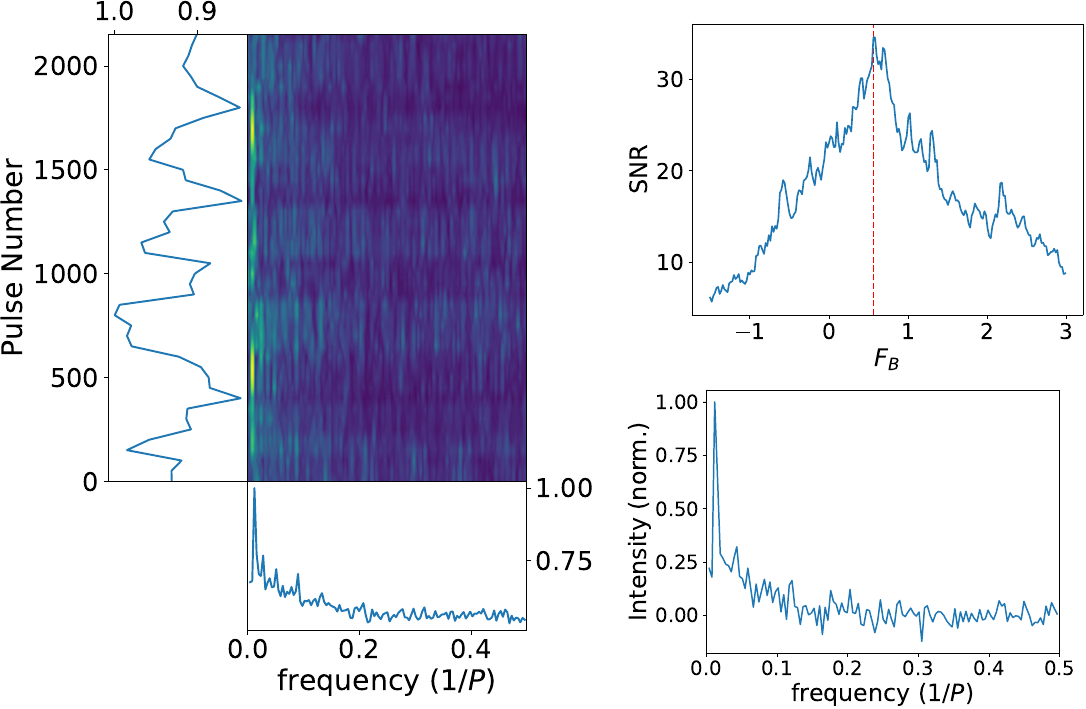}{0.46\textwidth}{PSR B1730$-$37}
         }
\caption{The left panel shows the time varying longitude-resolved fluctuation 
spectra (LRFS) estimated on the single pulse sequence. The 0/1 time series FFT 
is estimated for different cutoff levels and the signal to noise ratio (SNR) of
the periodic feature is shown in the top window of the right panel along with 
the maximum value (dashed vertical red line). The bottom window shows the 
average FFT of the 0/1 sequence with maximum SNR of periodic feature which
closely resembles the average LRFS spectra.
\label{fig:permod_1}}
\end{figure}

\begin{figure}
\gridline{\fig{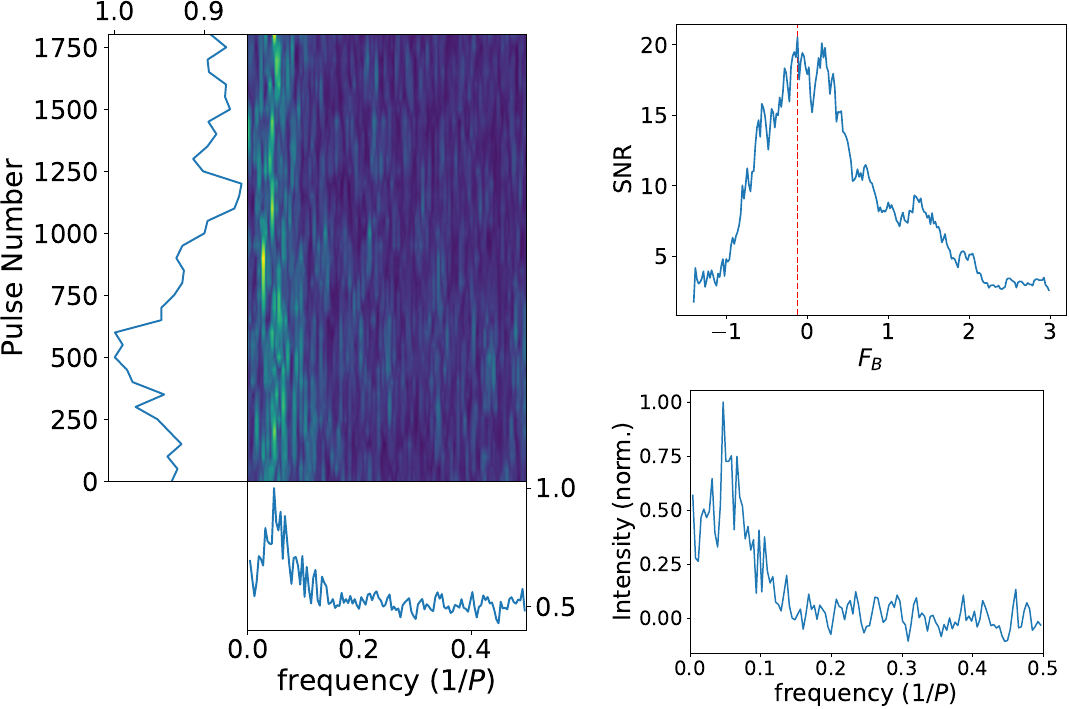}{0.46\textwidth}{PSR B1732$-$07}
	  \fig{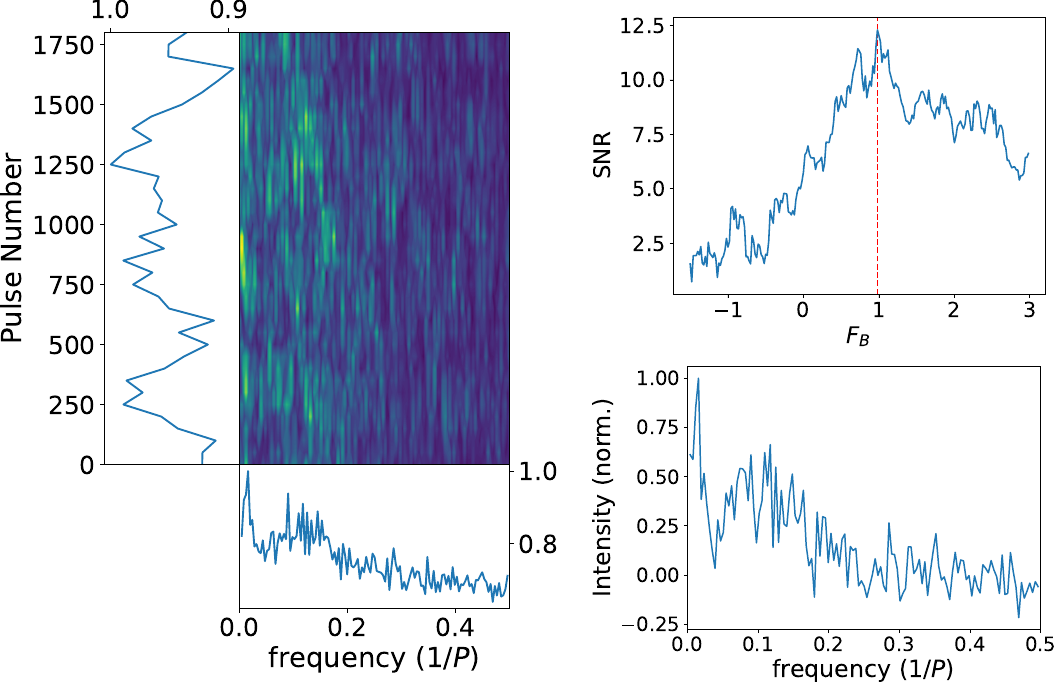}{0.46\textwidth}{PSR B1737$-$39}
	 }
\gridline{\fig{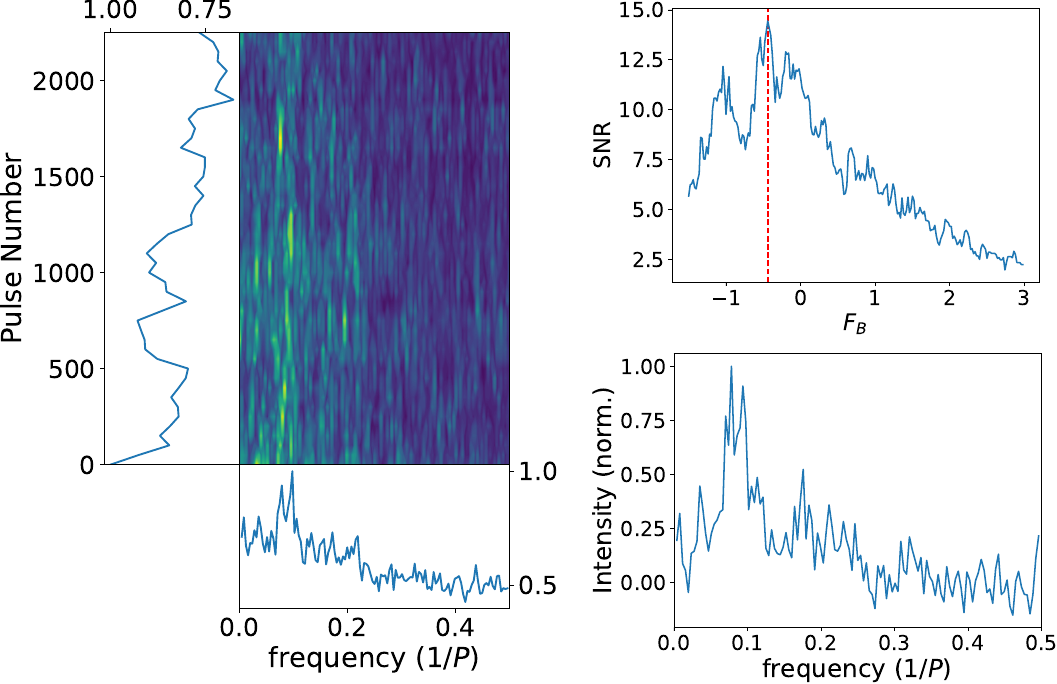}{0.46\textwidth}{PSR B1929+10}
	  \fig{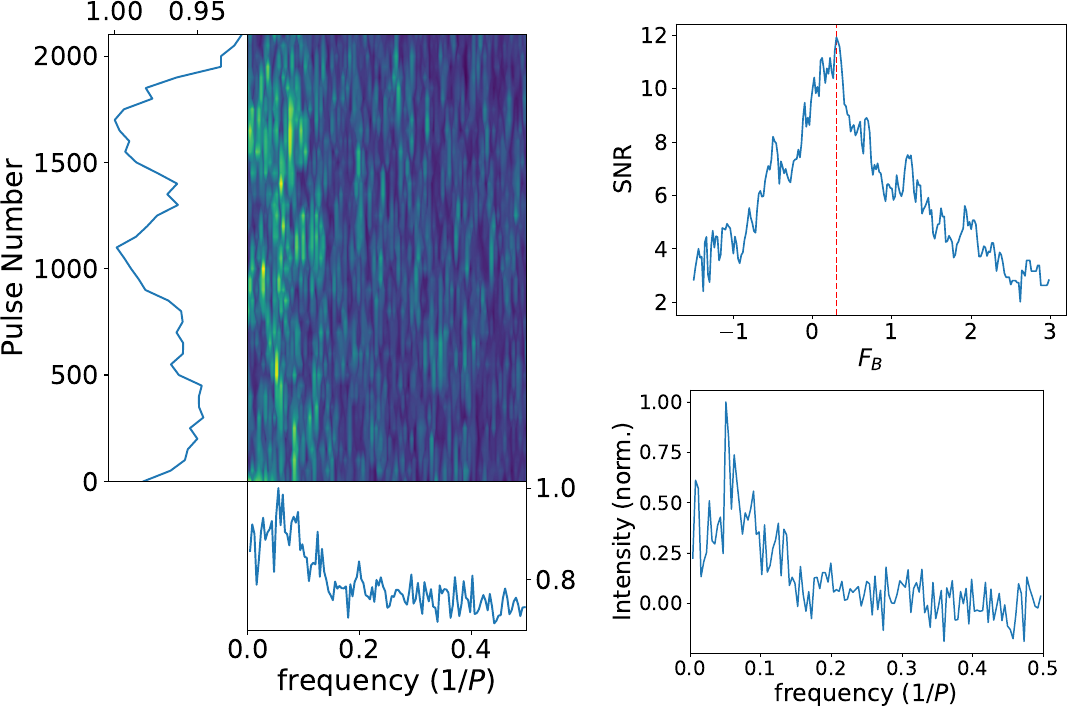}{0.46\textwidth}{PSR B1929+20}
         }
\gridline{\fig{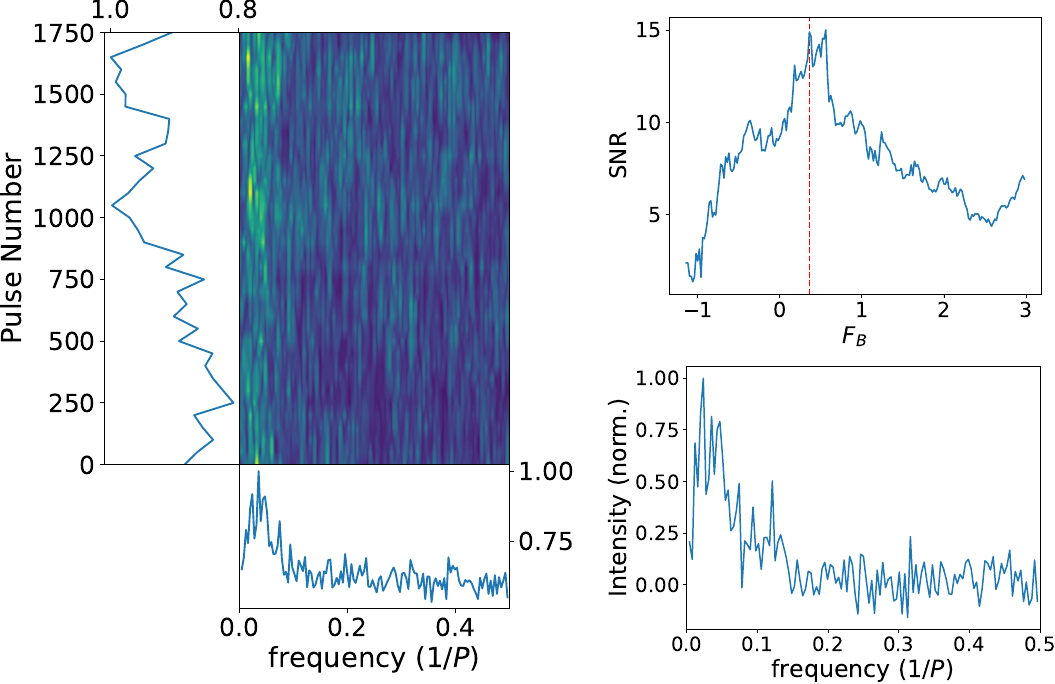}{0.46\textwidth}{PSR B2011+38}
	  \fig{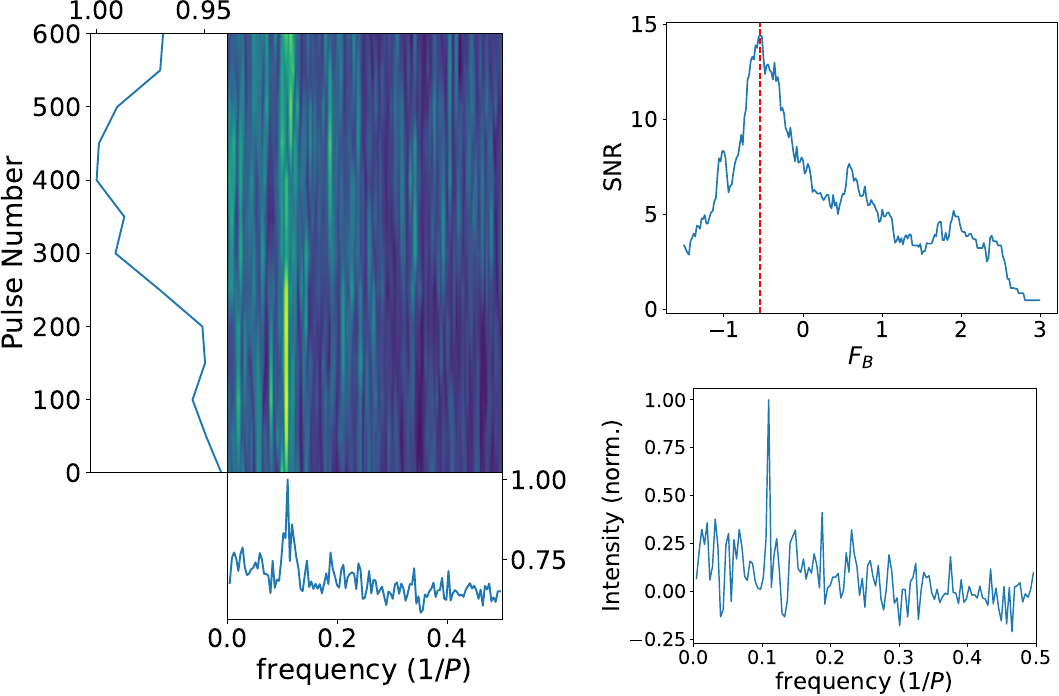}{0.46\textwidth}{PSR B2148+52}
	 }
\caption{The left panel shows the time varying longitude-resolved fluctuation
spectra (LRFS) estimated on the single pulse sequence. The 0/1 time series FFT
is estimated for different cutoff levels and the signal to noise ratio (SNR) of
the periodic feature is shown in the top window of the right panel along with
the maximum value (dashed vertical red line). The bottom window shows the
average FFT of the 0/1 sequence with maximum SNR of periodic feature which
closely resembles the average LRFS spectra.
\label{fig:permod_2}}
\end{figure}

\begin{figure}
\gridline{\fig{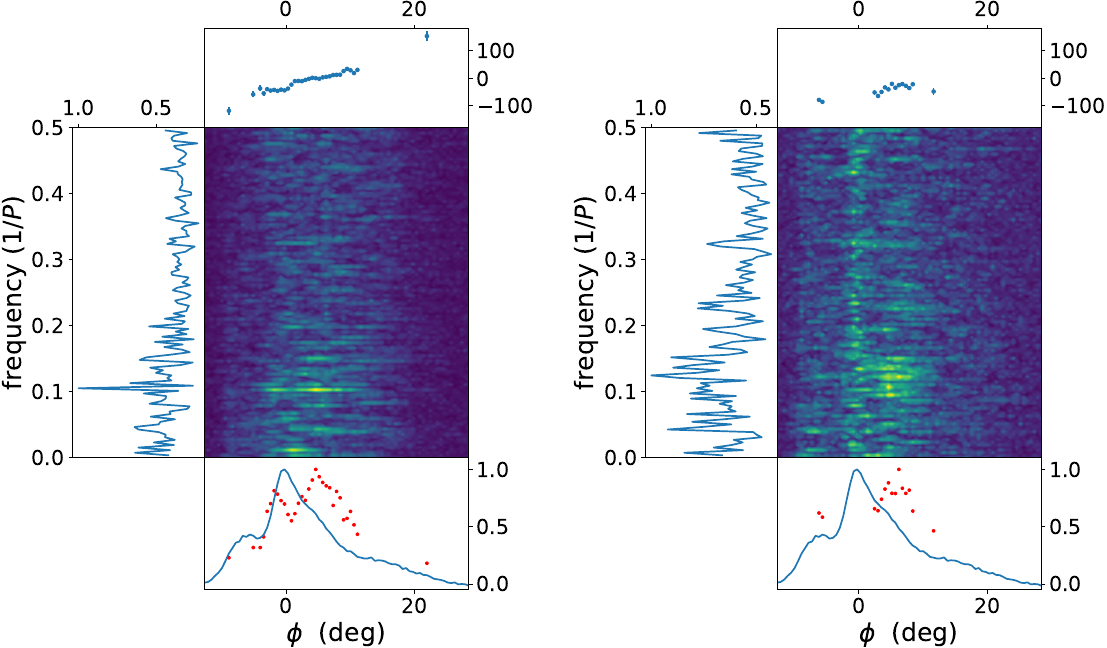}{0.45\textwidth}{PSR B0450+55}
	  \fig{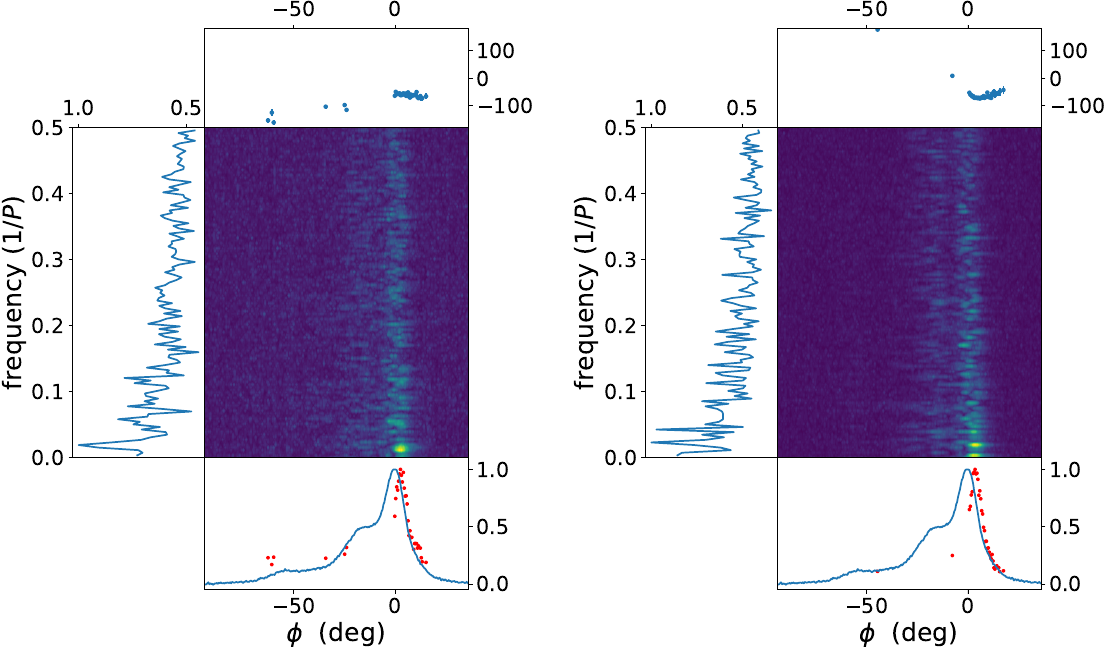}{0.45\textwidth}{PSR B0905$-$51}
         }
\gridline{\fig{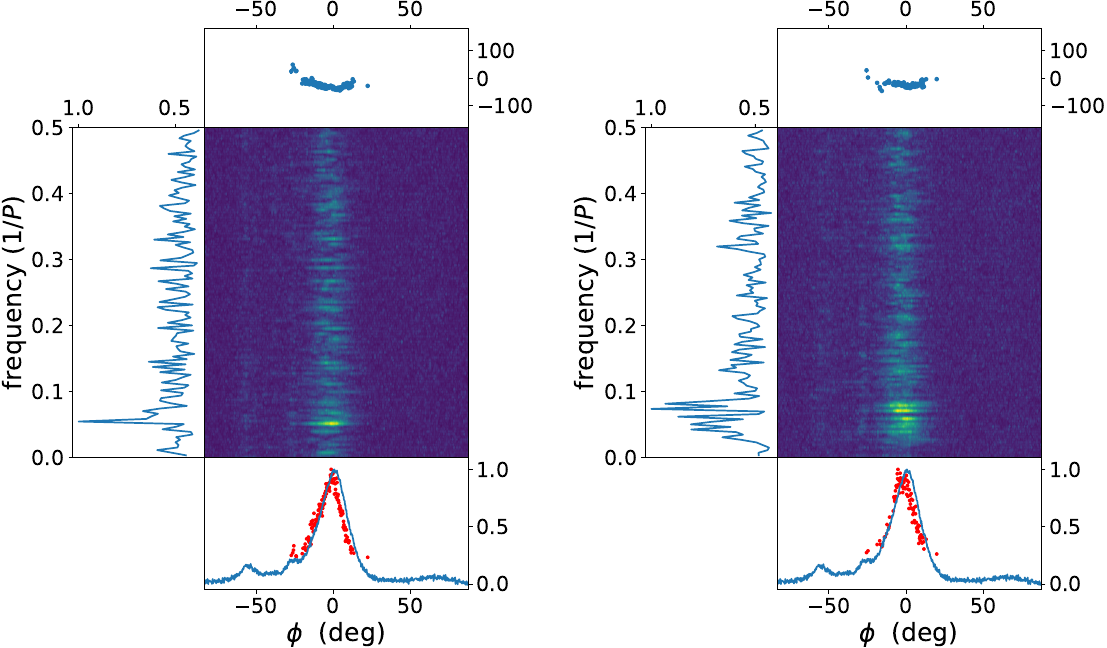}{0.45\textwidth}{PSR B1541+09}
          \fig{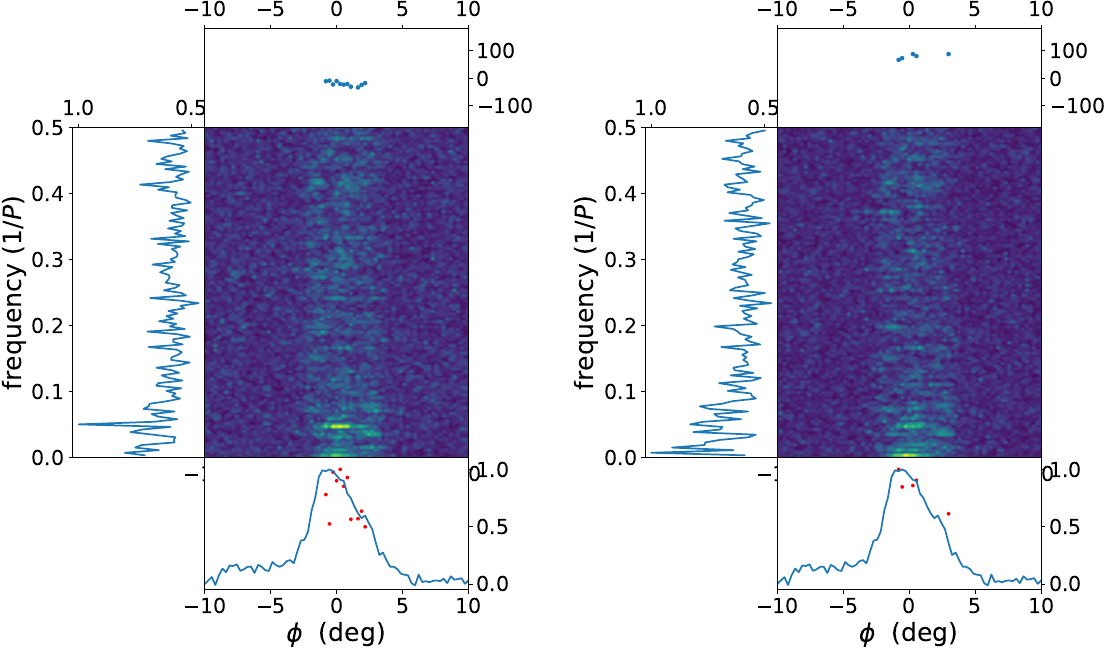}{0.45\textwidth}{PSR B1600$-$49}
         }
\gridline{\fig{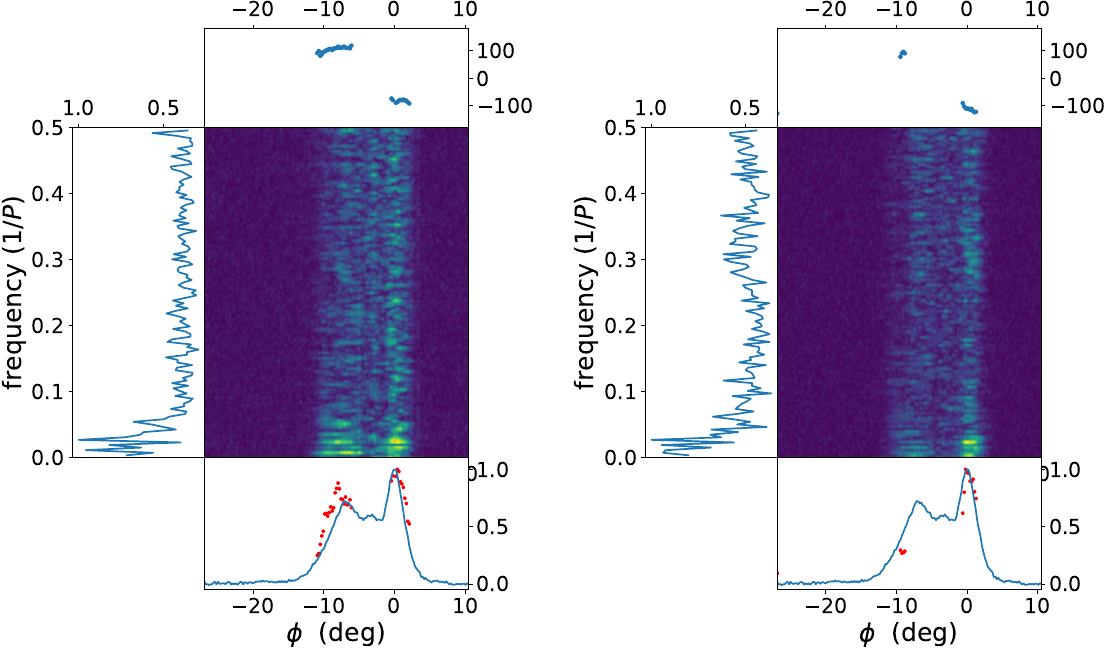}{0.45\textwidth}{PSR B1604$-$00}
          \fig{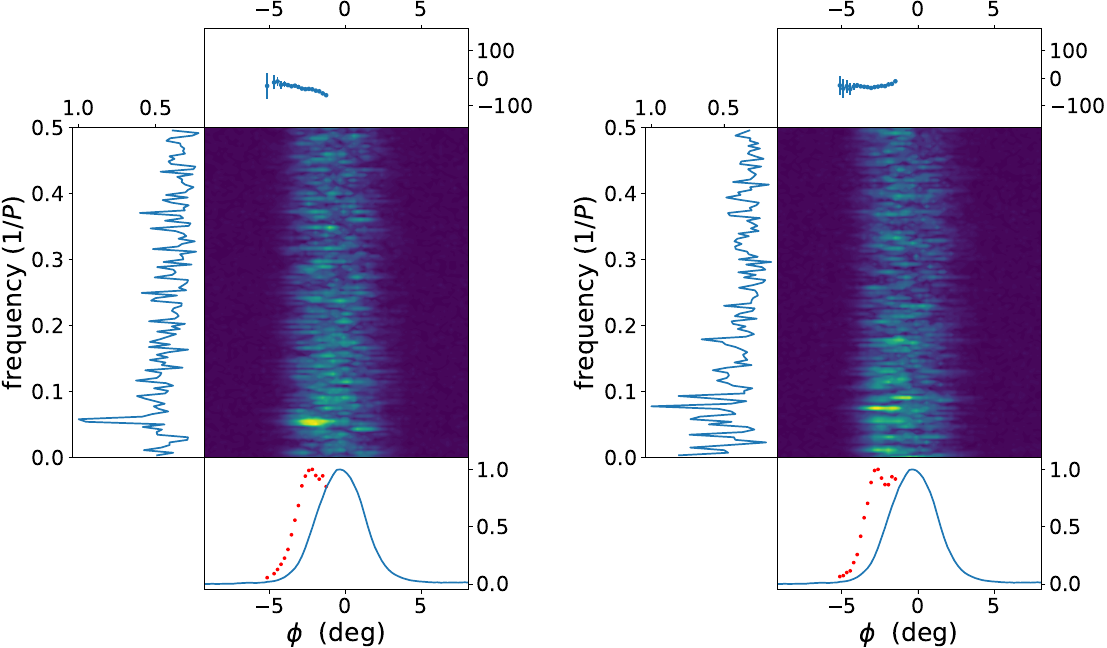}{0.45\textwidth}{PSR B1642$-$03}
         }
\gridline{\fig{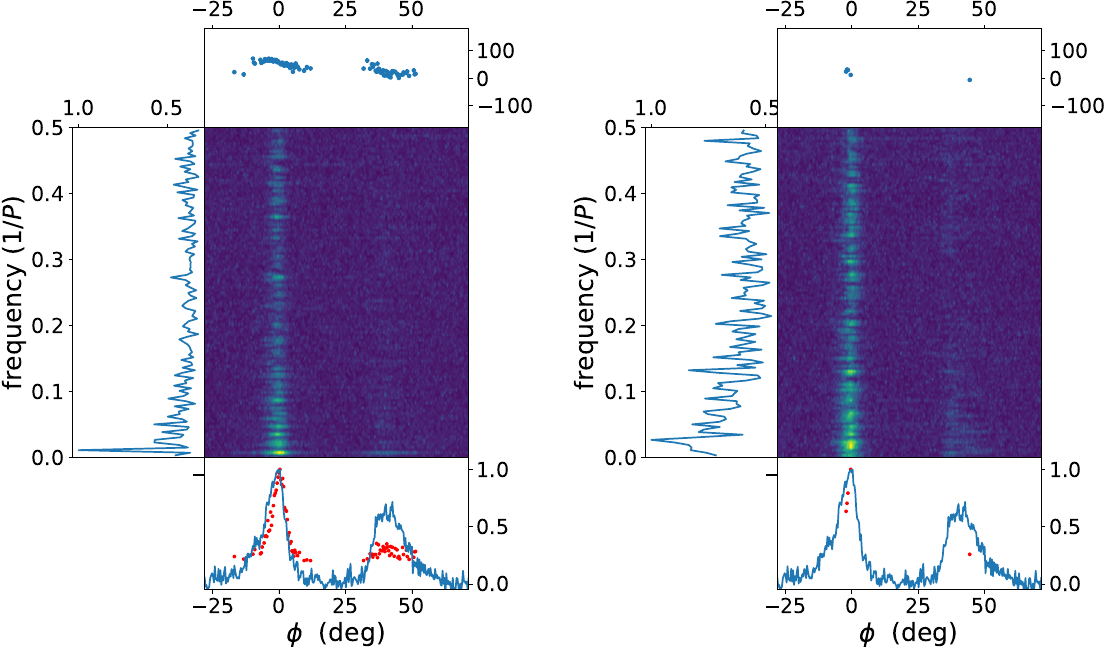}{0.45\textwidth}{PSR B1730$-$37}
          \fig{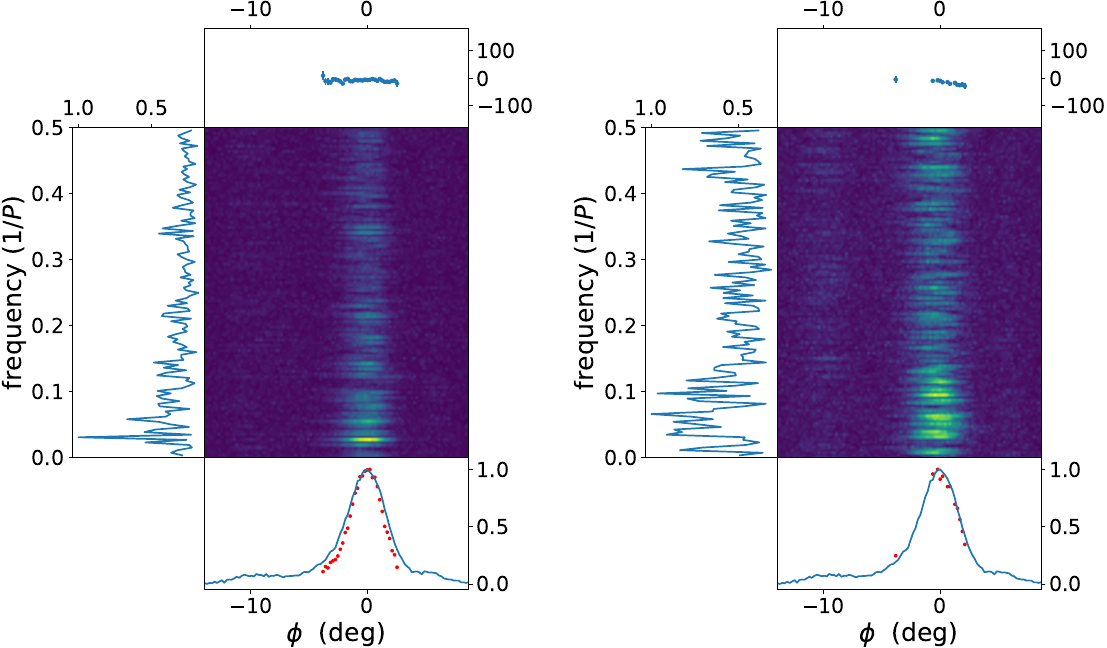}{0.45\textwidth}{PSR B1732$-$07}
         }
\caption{The periodic behaviour from different pulse sequence of the same 
pulsar. The left panel shows the LRFS with prominent periodic modulation while
the right panel shows diffuse structure.	
\label{fig:lrfs_1}}
\end{figure}

\begin{figure}
\gridline{\fig{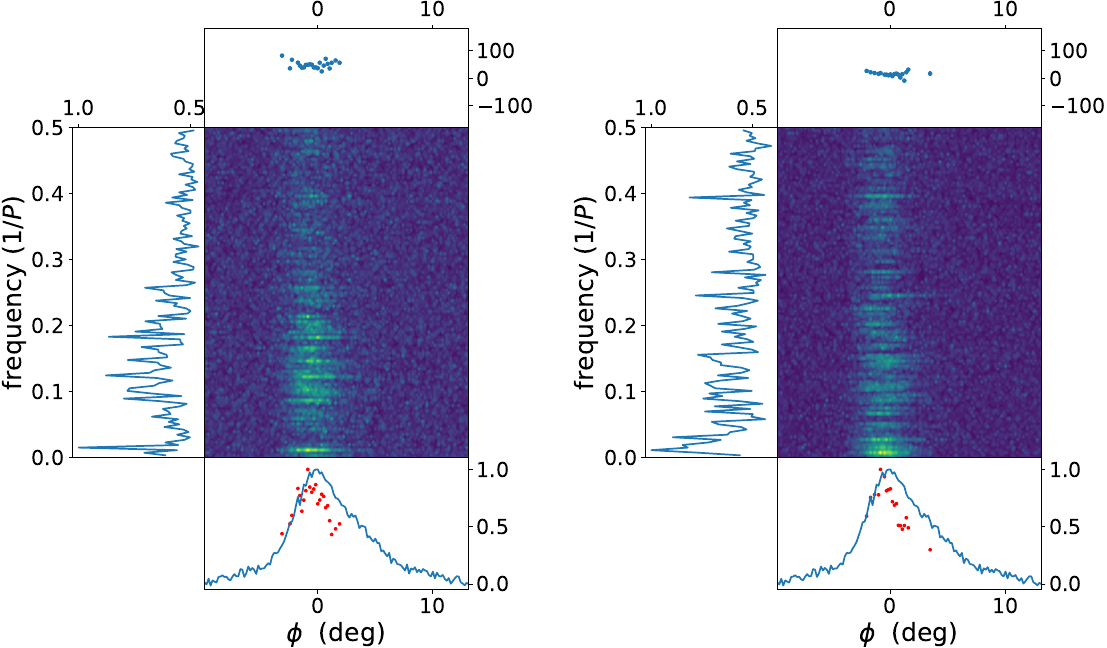}{0.45\textwidth}{PSR B1737$-$39}
          \fig{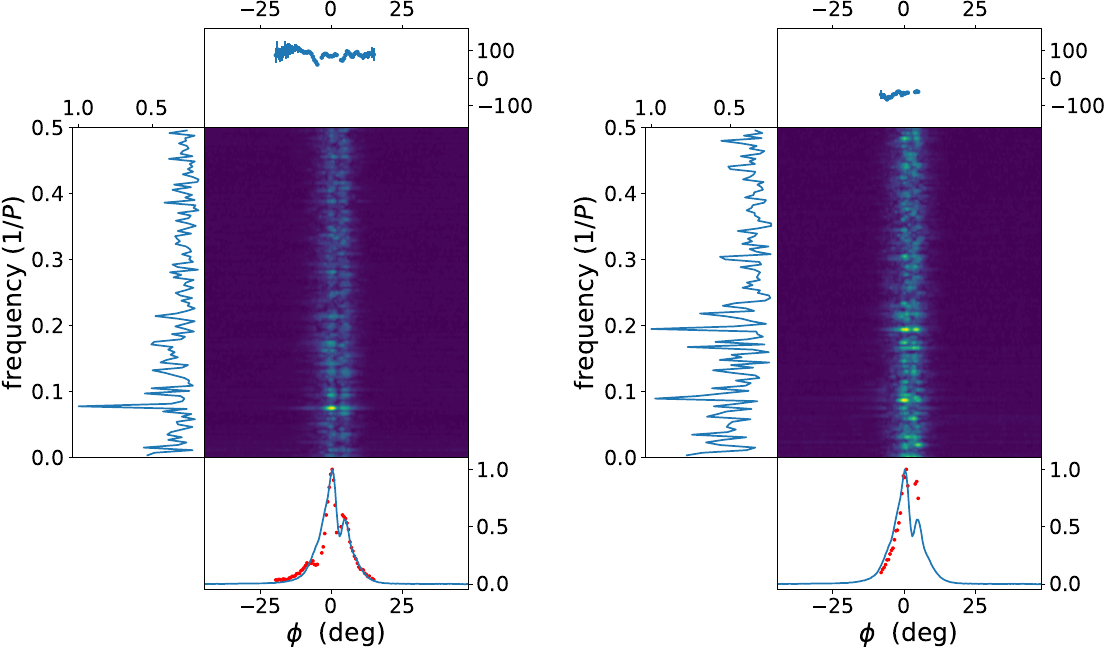}{0.45\textwidth}{PSR B1929+10}
         }
\gridline{\fig{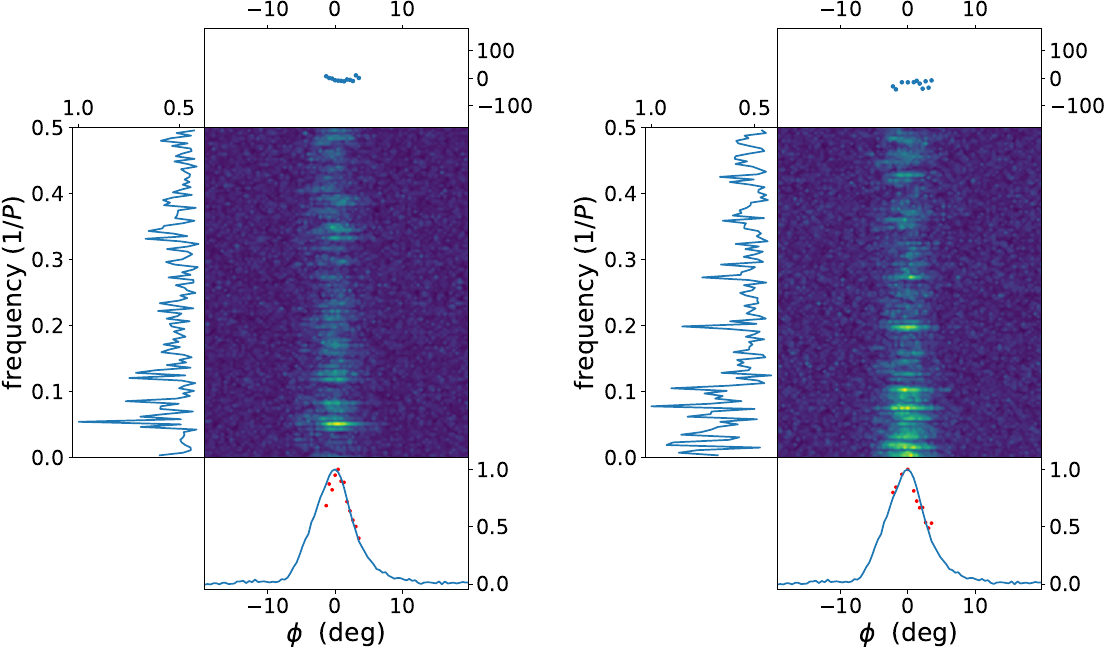}{0.45\textwidth}{PSR B1929+20}
          \fig{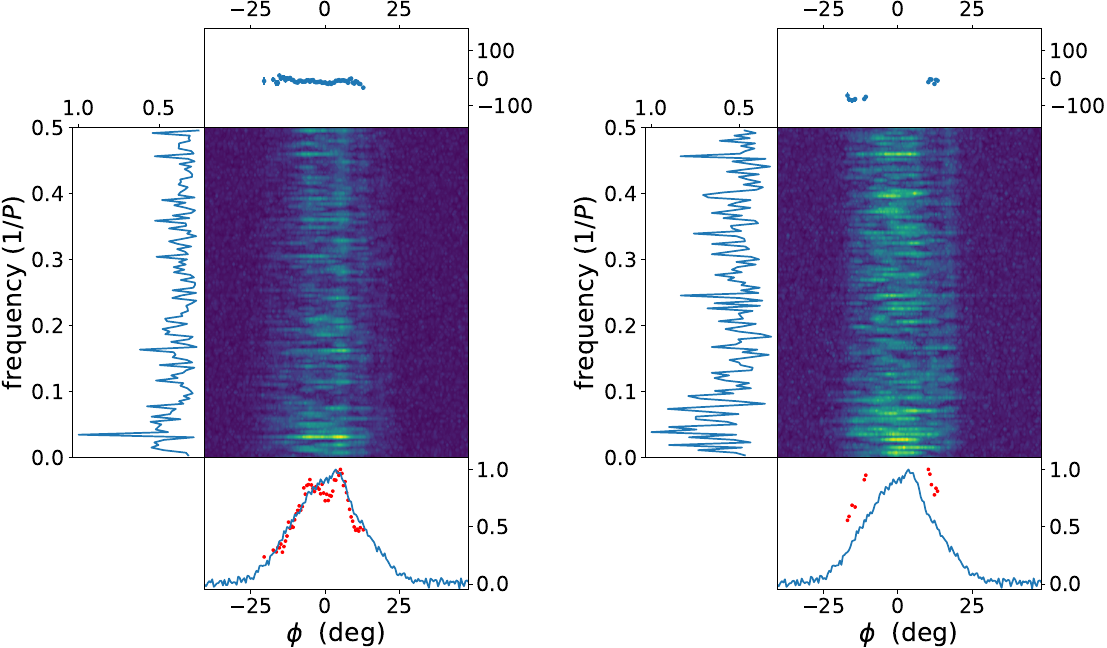}{0.45\textwidth}{PSR B2011+38}	
	 }
\gridline{\fig{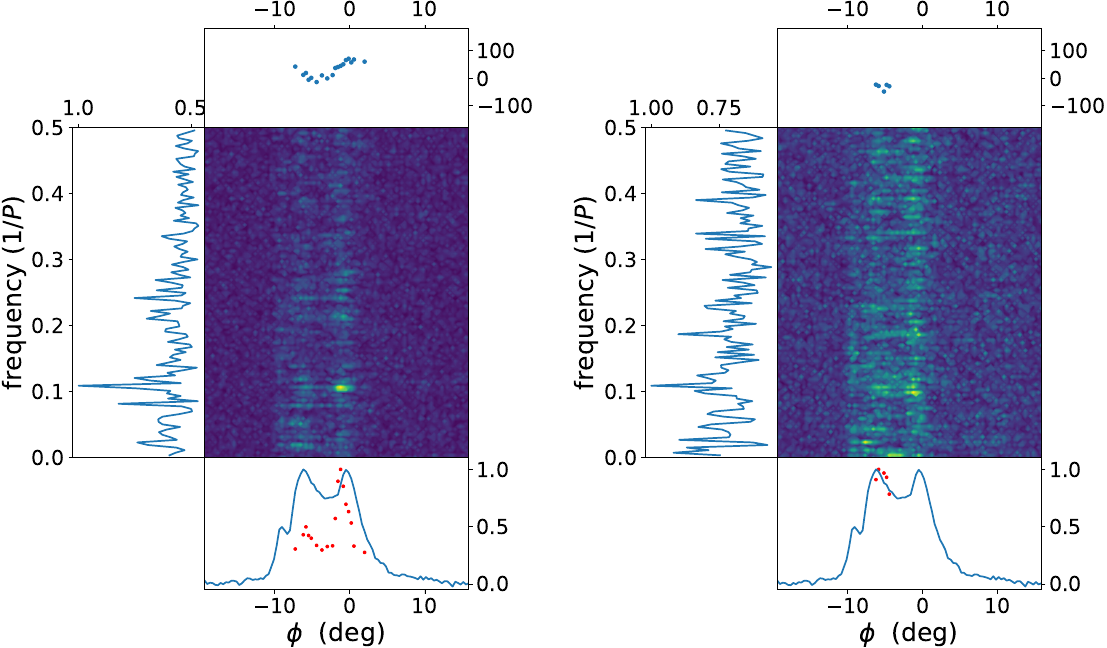}{0.45\textwidth}{PSR B2148+52}
	 }
\caption{The periodic behaviour from different pulse sequence of the same
pulsar. The left panel shows the LRFS with prominent periodic modulation while
the right panel shows diffuse structure.
\label{fig:lrfs_2}}
\end{figure}

\begin{figure}
\gridline{\fig{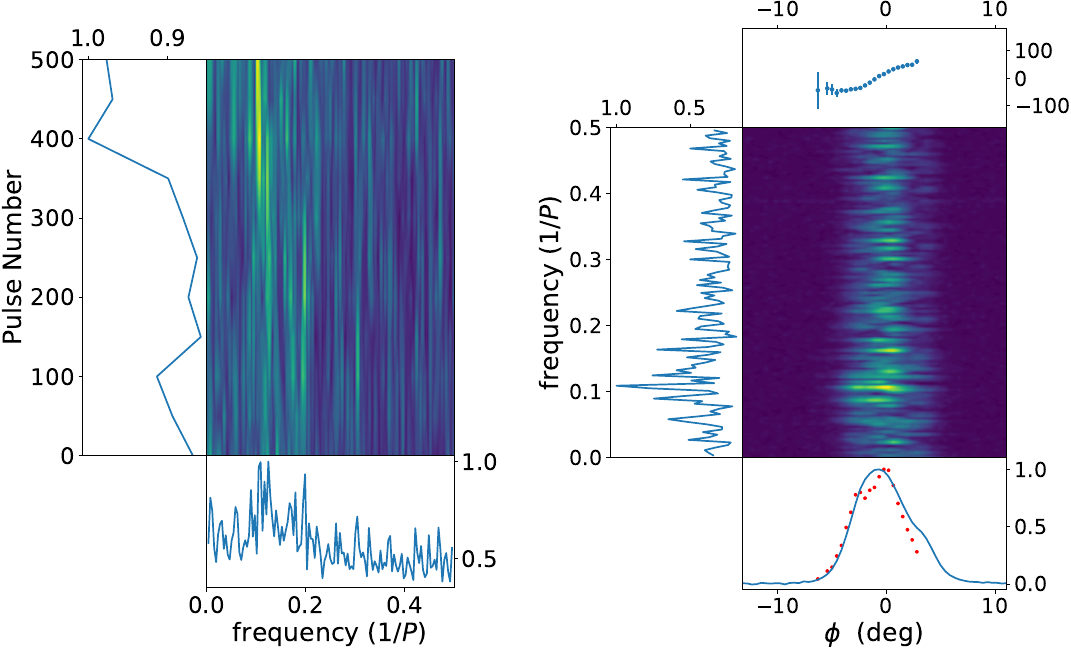}{0.45\textwidth}{PSR B0136+57}
          \fig{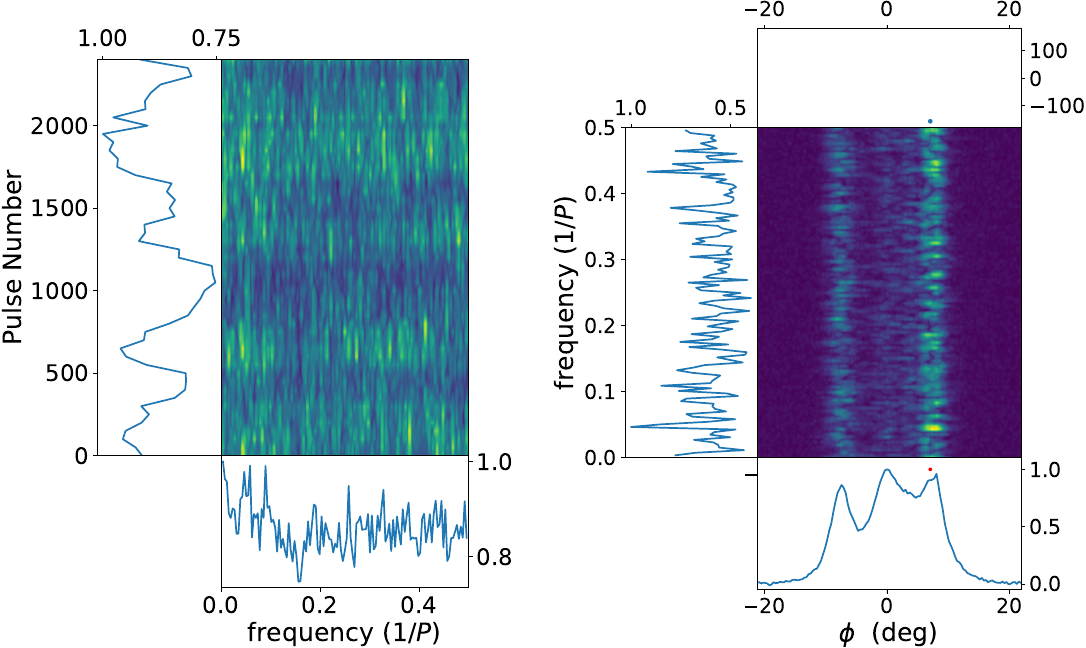}{0.45\textwidth}{PSR B0450$-$18}
         }
\gridline{\fig{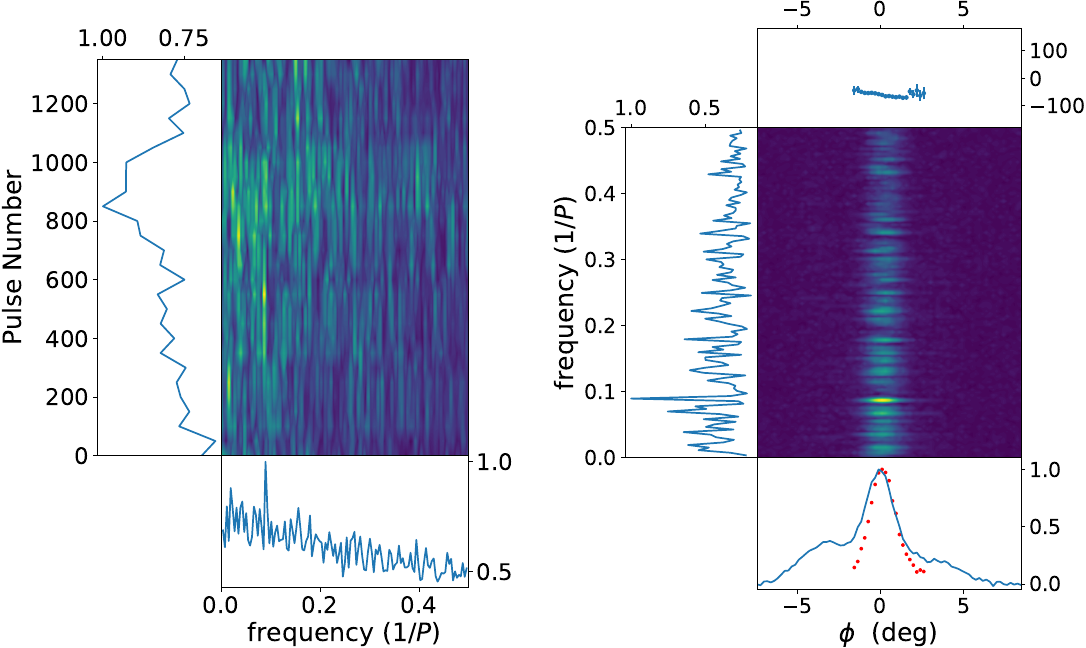}{0.45\textwidth}{PSR B1917+00}
          \fig{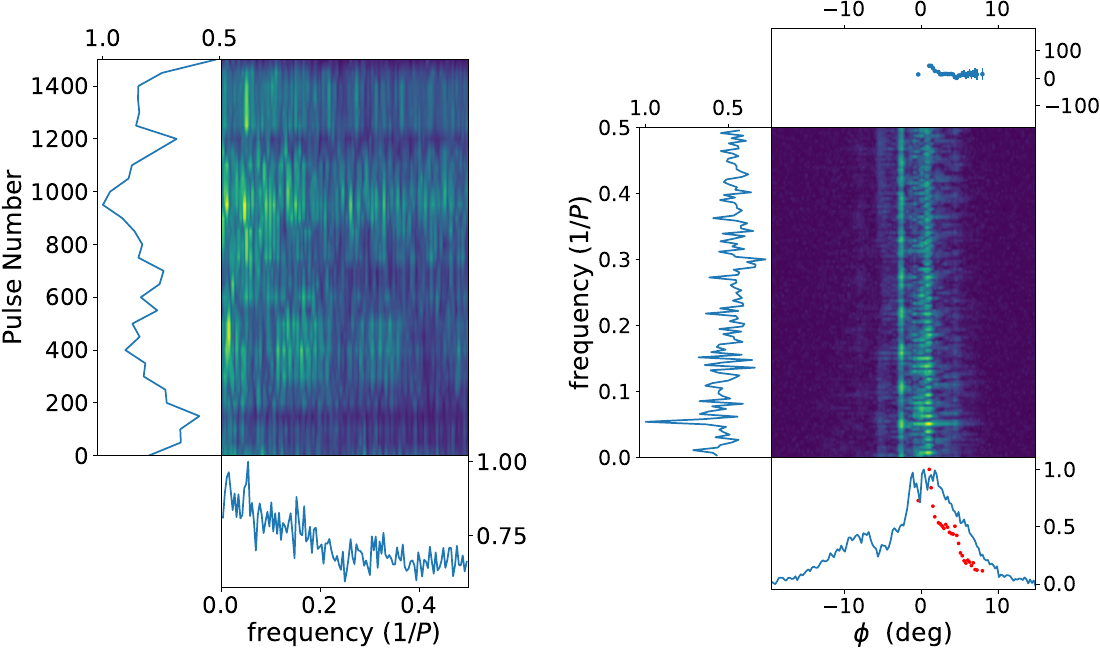}{0.45\textwidth}{PSR B2334+61}
         }
\caption{The left panel shows the time varying average LRFS for pulsars with
intermittent periodic modulation. The right panel shows the LRFS from a 
specific pulse sequence with more prominent periodic feature.
\label{fig:lrfs_int}}
\end{figure}

\appendix
The figures corresponding to the periodic modulation measurements of each 
pulsar are reported in the appendix. The time varying average LRFS and the 
average FFT obtained from 0/1 sequence corresponding to the two emission states
are shown in Fig.~\ref{fig:permod_1} and \ref{fig:permod_2}. The LRFS from two 
specific pulse sequence highlighting the variation in the modulation behaviour
of pulsars with phase stationary modulation and periodic modulations with phase
shifts are shown in Fig.~\ref{fig:lrfs_1} and \ref{fig:lrfs_2}. The time 
varying LRFS and an example of periodic behaviour in the LRFS for the four
pulsars with intermittent modulations is shown in Fig.~\ref{fig:lrfs_int}.

\end{document}